\documentclass[final]{siamltex}

\usepackage{amssymb}
\usepackage{amsmath}
\usepackage{color}
\usepackage{graphicx}
\usepackage{bm}
\usepackage{accents}

\usepackage{datetime}

\usepackage[colorlinks=true, linkcolor=blue, citecolor=blue, filecolor=blue,
runcolor=blue, urlcolor = blue]{hyperref}

\usepackage[ruled,vlined]{algorithm2e}

\newcommand{\R}{\mathbb{R}}

\newcommand{\covG}{\mathcal{G}}

\renewcommand{\phi}{\varphi}
\renewcommand{\epsilon}{\varepsilon}

\newcommand{\bX}{\mathbf{X}}

\newcommand{\A}{\mathbb{A}}

\newcommand{\mb}[1]{\mathbf{#1}}

\newcommand{\OpFSCouple}{\Gamma}
\newcommand{\OpSFCouple}{\Lambda}

\newcommand{\Eng}{\Phi}

\newcommand{\Fentrain}{\bm{f}_{s}}

\newcommand{\OpAdjoint}{{\dagger}}

\newcommand{\subtxt}[1]{ {\mbox{\tiny #1}} }

\definecolor{issuePJA_color}{rgb}{1.0,0.0,0.8}

\definecolor{commentPJA_color}{rgb}{1.0,0.0,0.8}

\definecolor{issuePP_color}{rgb}{0.1,1.0,0.1}

\definecolor{commentPP_color}{rgb}{0.1,1.0,0.1}

\DeclareGraphicsExtensions{.pdf}
\DeclareGraphicsExtensions{.png}

\begin{document}

\title{Spatially Adaptive Stochastic Methods for Fluid-Structure Interactions Subject to Thermal Fluctuations in Domains with Complex Geometries}

\author{Pat Plunkett
\thanks{University of California, 
Department of Mathematics}
\and
Jon Hu
\thanks{DOE Sandia} 
\and
Chris Siefert
\thanks{DOE Sandia, Sandia National Laboratories is a multiprogram laboratory managed and operated by Sandia Corporation, a wholly owned subsidiary of Lockheed Martin Corporation, for the U.S. Department of Energy's National Nuclear Security Administration under contract DE-AC04-94AL85000.} 
\and
Paul J. Atzberger
\thanks{University of California, 
Department of Mathematics , Santa Barbara, CA 93106; 
e-mail: atzberg@math.ucsb.edu; phone: 805-893-3239;
Work supported by DOE CM4 and NSF CAREER Grant DMS-0956210.}
}

\maketitle

\begin{abstract}
We develop stochastic mixed finite element methods for spatially adaptive simulations of fluid-structure interactions when subject to thermal fluctuations.  To account for thermal fluctuations, we introduce a discrete fluctuation-dissipation balance condition to develop compatible stochastic driving fields for our discretization.  We perform analysis that shows our condition is sufficient to ensure results consistent with statistical mechanics.  We show the Gibbs-Boltzmann distribution is invariant under the stochastic dynamics of the semi-discretization.  To generate efficiently the required stochastic driving fields, we develop a Gibbs sampler based on iterative methods and multigrid to generate fields with $O(N)$ computational complexity.  Our stochastic methods provide an alternative to uniform discretizations on periodic domains that rely on Fast Fourier Transforms.   To demonstrate in practice our stochastic computational methods, we investigate within channel geometries having internal obstacles and no-slip walls how the mobility/diffusivity of particles depends on location.  Our methods extend the applicability of fluctuating hydrodynamic approaches by allowing for spatially adaptive resolution of the mechanics and for domains that have complex geometries relevant in many applications.
\end{abstract}


\pagestyle{myheadings}
\thispagestyle{plain}
\markboth{P. PLUNKETT, P.J. ATZBERGER}
{STOCHASTIC MULTIGRID FOR BROWNIAN-STOKESIAN DYNAMICS}

\section*{Introduction} 
We develop general computational methods for applications involving the microscopic mechanics of spatially extended elastic bodies within a fluid that are subjected to thermal fluctuations.  Motivating applications include the study of the microstructures of complex fluids~\cite{Doi1986}, lipid bilayer membranes~\cite{Helfrich1973,Lipowsky1991, AtzbergerThermoStat2013}, and micro-mechanical devices~\cite{Raman2008,Ho1998}. Even in the deterministic setting, the mechanics of fluid-structure interactions pose a number of difficult and long-standing challenges owing to the rich behaviors that can arise from the interplay of the fluid flow and elastic stresses of the microstructures~\cite{Dowell2001,Shelley2011}.  To obtain descriptions tractable for analysis and simulations, approximations are often introduced into the fluid-structure coupling.  For deterministic systems, many spatially adaptive numerical methods have been developed for approximate fluid-structure interactions~\cite{Griffith2009,
Minion1996,Griffith2007,Roma1999,Almgren1998,Bell1996}.  In the presence of thermal fluctuations, additional challenges arise from the need to capture in computational methods the appropriate propagation of fluctuations throughout the discretized system to obtain results consistent with statistical mechanics.  In practice, challenges arise from the very different dissipative properties of the discrete operators relative to their continuum differential counterparts.  These issues have important implications for how stochastic fluctuations should be handled in the discrete setting.   Even when it is possible to formulate stochastic driving fields in a well-founded manner consistent with statistical mechanics, these Gaussian random fields have often many degrees of freedom and non-trivial spatial correlations that can be difficult to sample without significant computational expense.  Many finite difference methods on uniform periodic meshes have been developed for fluctuating hydrodynamics~\cite{Bell2007,
DeFabritiis2007,Sharma2004a,
Donev2010Accuracy,
AtzbergerSIB2007,AtzbergerChannel2013,AtzbergerSELM2011}.
One of the main reasons that fluctuating hydrodynamics is treated on uniform periodic domains is so that stochastic driving fields can be generated using Fast Fourier Transforms (FFTs)~\cite{AtzbergerSIB2007,AtzbergerSELM2011}.  
Here, we take a different approach by developing stochastic methods based on Finite Element Methods for fluctuating hydrodynamics and provide an alternative to Fast Fourier Transforms for the generation of stochastic driving fields.  Our approach allows for non-uniform spatially adaptive discretizations on non-periodic domains with geometries more naturally encountered in many applications.

We develop Finite Element Methods with properties that facilitate the introduction of stochastic driving fields and their efficient generation.  We show our discretization approach provides operators that satisfy certain symmetry and commutation conditions that are important when subject to the incompressibility constraint for how thermal fluctuations propagate throughout the discrete system.  We formulate the stochastic equations for our fluid-structure system subject to thermal fluctuations in Section~\ref{sec_fluid_struct}.  We introduce for a given spatial discretization our general procedure for deriving compatible stochastic driving fields that model the thermal fluctuations in a manner consistent with statistical mechanics in Section~\ref{sec_FEM_discr}.  To obtain the stochastic driving fields with the required spatial correlation structure, we develop stochastic iterative methods based on multigrid to generate the 
Gaussian random fields with computational complexity $O(N)$ in Section~\ref{sec_stoch_field_gen}.  We present validation of our stochastic numerical methods with respect to the hydrodynamic coupling and thermal fluctuations in Section~\ref{sec_validation}.  To demonstrate our approach in practice, we present simulations of a few example systems in Section~\ref{sec_application}.   

Overall, our approach extends the range of problems that can be treated numerically with fluctuating hydrodynamic methods by allowing for arbitrary geometries with walls having no-slip boundary conditions and by allowing for spatially adaptive resolution.  Many of the central ideas used for our numerical approximation of the fluctuating hydrodynamic equations should also be applicable in the approximation of other parabolic Stochastic Partial Differential Equations (SPDEs).   We expect our stochastic numerical methods for fluctuating hydrodynamics to be useful in applications where the domain geometry plays an important role.

\section{Fluid-Structure Hydrodynamics and Fluid-Structure Interactions} 
\label{sec_fluid_struct}
We describe the mechanics of fluid-structure interactions subject to thermal fluctuations using the Stochastic Eulerian Lagrangian Method (SELM)~\cite{AtzbergerSELM2011}.  In the inertial regime this is given by momentum equations for the fluid coupled to momentum equations for the microstructures~\cite{AtzbergerSELM2011}.  We consider here the regime in which the fluid-structure coupling is strong and the microstructures are mass density matched with the fluid~\cite{AtzbergerSELM2011,AtzbergerTabak2013}.  This regime is closely related to the Stochastic Immersed Boundary Method~\cite{AtzbergerSIB2007,AtzbergerTabak2013,Peskin2002,Bringley2008}.  In this regime, we use the time-dependent Stokes equations for the fluid coupled to an equation of motion for the microstructures
\begin{equation}
  \label{eq:TDStokes}
  \begin{aligned}
    \rho\frac{\partial \bm{u}}{\partial t} &= \mu\Delta\bm{u} - \nabla p +
\Fentrain + \bm{f}_{thm} &\text{in }\Omega\\
    \nabla\cdot\bm{u} &= 0 &\text{in }\Omega\\
    \bm{u}|_{\partial\Omega} &= 0.
  \end{aligned}
\end{equation}
The elastic microstructures with configuration $\mb{X}$ are given by the following equation of motion and coupling condition that models the  bidirectional coupling between the fluid and microstructures
\begin{eqnarray}  
\label{equ_dX}
\frac{d\mathbf{X}}{dt} & = & \Gamma \mathbf{u} \\
\label{equ_f_int}
\Fentrain           & = & \Lambda\left\lbrack-\nabla
\Phi(\mb{X})\right\rbrack.
\end{eqnarray}
The thermal fluctuations are taken into account by the Gaussian random field $\bm{f}_{thm}$ which when decomposed into a mean and fluctuating part $\bm{f}_{thm} = \bar{\bm{f}}_{thm} + \tilde{\bm{f}}_{thm}$ has the form 
\begin{eqnarray}  
\label{equ_thermal_drift}
\bar{\bm{f}}_{thm}   & = & \langle \bm{f}_{thm} \rangle = k_B{T}\nabla_{\bX} \cdot \Lambda \\
\label{equ_cov_fluid}
\langle \tilde{\bm{f}}_{thm}(\mathbf{x}) \tilde{\bm{f}}_{thm}^T(\mathbf{y}) \rangle & = & 2\mu \Delta C(\mathbf{x} - \mathbf{y}) \\ 
C(\mathbf{x} - \mathbf{y}) & = & k_B{T} \rho^{-1} \delta(\mathbf{x} - \mathbf{y}).
\end{eqnarray}  
These stochastic driving fields were derived for the mechanical system using the SELM framework in~\cite{AtzbergerSELM2011}.  A notable difference with the original formulation of the Stochastic Immersed Boundary Method (SIBM) is the presence of the thermal drift term in equation~\ref{equ_thermal_drift} which arises from the more systematic treatment through stochastc averaging to obtain in this regime the stochastic fluid-structure equations which handles the generalised configuration-momentum coordinates used in such descriptions~\cite{AtzbergerSELM2011,AtzbergerTabak2013}.

In the notation, the $\mathbf{u}$ is the fluid velocity field, $\mathbf{X}$ is the collective microstructure configuration, $p$ the pressure, and $\mu$ the dynamic shear viscosity. The hydrodynamic equations~\ref{eq:TDStokes} account for the microstructure interaction through $\Fentrain$. For short, we let $\mathbf{f} =\Fentrain + \bm{f}_{thm}$. In the motion of the microstructures in response to the fluid flow is given by equation~\ref{equ_dX}.  Similar to the Immersed Boundary Method~\cite{Peskin2002,Bringley2008}, the operator $\Gamma$ provides a model for how the microstructure locally responds to the fluid flow. The influence that the microstructure has on the nearby fluid is given by equation~\ref{equ_f_int}. The $\Lambda$ operator models the neighborhood of surrounding fluid that is affected by forces acting on the microstructure.  These operators can be chosen quite generally provided they satisfy the adjoint condition~\cite{Peskin2002,Bringley2008,AtzbergerSELM2011}
\begin{eqnarray}
\label{equ_adjoint}
\langle \Lambda \mb{V}, \mb{u} \rangle_{\Omega} = \langle \mb{V}, \Gamma \mb{u} \rangle_{M}
\end{eqnarray}
where $\mb{u}$ has the same form as the fluid velocity field and $\mb{V}$ the same form as the microstructure velocity.  The $\langle \mb{f},\mb{g} \rangle_{\Omega} = \int_{\Omega} \mb{f}(x)\cdot\mb{g}(x) dx$ denotes integration over the spatial domain of the fluid  and $\langle \mb{F}, \mb{G} \rangle_{M} = \sum \mb{F}_j\cdot\mb{G}_j$ denotes the dot-product over the microstructure degrees of freedom.  With these assumptions, the equations~\ref{eq:TDStokes}--~\ref{equ_f_int} describe the mechanics of the fluid-structure interactions subject to thermal fluctuations in the physical regime where the microstructure is strongly 
coupled to the fluid and mass density-matched with the fluid~\cite{AtzbergerSELM2011, AtzbergerTabak2013}.

In the regime where the hydrodynamics relaxes rapidly relative to the time-scale of microstructure motions, the fluid-structure dynamics can be further reduced~\cite{AtzbergerSELM2011, AtzbergerTabak2013} to obtain
\begin{eqnarray}  
\label{equ_SELM_stokes}
\frac{d\mathbf{X}}{dt} & = & H \left\lbrack-\nabla \Phi(\mb(X))\right\rbrack +
k_B{T}\nabla\cdot H + \mb{h} \\
\langle \mb{h}\mb{h}^T \rangle & = & 2k_B{T} H.
\end{eqnarray}
We refer to this as the overdamped Stokes regime which is similar to Brownian-Stokesian dynamics~\cite{Brady1988, ErmakMcCammon1978}. The effective hydrodynamic coupling tensor of the microstructure is given by $H = \Gamma \wp \mathcal{L} \wp^T \Lambda$ with $\mathcal{L} = \mu^{-1} \Delta^{-1}$ and $\wp = \mathcal{I} - \nabla \Delta^{-1} \nabla \cdot$ is the solenoidal projection operator imposing the hydrodynamic incompressibility~\cite{Chorin1968}. The thermal fluctuations are taken into account through the stochastic driving term $\mb{h}$ and the thermal drift-divergence term $k_B{T}\nabla\cdot H$.  The above equations can be shown to have the Gibbs-Boltzmann distribution invariant with detailed-balance under the stochastic dynamics~\cite{AtzbergerSELM2011, AtzbergerTabak2013}.
  
\section{Finite Element Semi-Discretization : Mixed-Method}
\label{sec_FEM_discr}
We develop mixed finite element methods for semi-discretization of the fluid-structure system where the fluid is governed by the 
time-dependent Stokes equations.  We perform a stochastic reduction of the semi-discretized equations in the limit where the hydrodynamics is assumed to relax rapidly on the time-scale of the motions of the microstructures.  This reduction provides a numerical discretization of the fluid-structure system in the overdamped regime.

In the weak formulation of the fluid equations, we consider the fluid velocity $\bm{u}$ in the space $H_0^1(\Omega)^3$ and the pressure $p$ in the space $L^2(\Omega)$.  The 
$H_0^1(\Omega)^3$ denotes the triple product of the Sobolev space under the $L^2$-norm with weak derivatives up to order one and zero trace on the domain boundary~\cite{Brenner2008}.  
The weak formulation of the fluid equations is
\begin{equation}
  \label{eq:VStokes}
  \begin{aligned}
    \rho \langle\partial_t\bm{u},\bm{\phi}\rangle &= -\mu a(\bm{u},\bm{\phi}) +
b(\bm{\phi},p) + \langle\bm{f},\bm{\phi}\rangle, && \text{for all }\bm{\phi}\in
H_0^1(\Omega),\\
    b(\bm{u},\psi) &= 0, && \text{for all }\psi\in L^2(\Omega).
  \end{aligned}
\end{equation}
The $\langle\cdot,\cdot\rangle$ denotes the $L^2$-inner product.
The $a:H_0^1(\Omega)^3\times H_0^1(\Omega)^3\rightarrow \R$ and $b:H_0^1(\Omega)^3\times L^2(\Omega)\rightarrow \R$ are the continuous bilinear forms 
\begin{align*}
  a(\bm{u},\bm{v}) &= \langle\nabla\bm{u},\nabla\bm{v}\rangle =
\int_{\Omega}\nabla\bm{u}:\nabla\bm{v}~d\bm{x},\\
  b(\bm{u},q) &= \langle q,\nabla\cdot\bm{u}\rangle =
\int_{\Omega}q\nabla\cdot\bm{u}~d\bm{x}.
\end{align*}
We use the Ritz-Galerkin approximation of the variational problem corresponding to restriction to the finite dimensional subspaces $\mathcal{V}_h\subset H_0^1(\Omega)^3$ and $\mathcal{P}_h\subset L^2(\Omega)$.  Our specific choice of spaces $\mathcal{V}_h,\mathcal{P}_h$ will be discussed in Section~\ref{sec_element_choice}.  The exact solution to the variational problem in equation~\ref{eq:VStokes} is approximated by $\bm{u}_h\in \mathcal{V}_h$ and 
$p_h\in \mathcal{P}_h$ satisfying the finite-dimensional problem 
\begin{equation}
  \label{eq:FEMStokes}
  \begin{aligned}
    \rho\langle\partial_t\bm{u}_h,\bm{\phi}_i\rangle &= -\mu
a(\bm{u}_h,\nabla\bm{\phi}_i) + b(\bm{\phi}_i,p_h) +
\langle\bm{f},\bm{\phi}_i\rangle, && \text{for all }\bm{\phi}_i\in
\mathcal{V}_h\\
    b(\bm{u}_h,\psi_i) &= 0, && \text{for all }\psi_i\in \mathcal{P}_h.
  \end{aligned}
\end{equation}
This can be expressed in matrix form as 
\begin{equation}
  \label{eq:DiscreteStokes}
  \begin{aligned}
    \rho M\dot{U} &= -\mu LU + GP + F\\
    DU &= 0.
  \end{aligned}
\end{equation}
We represent functions using a basis $\mathcal{V}_h = \text{span}\{\bm{\phi}_i\}$ and a basis $\mathcal{P}_h = \text{span}\{\psi_i\}$ with
\begin{displaymath}
  \bm{u}_h = \sum_iU_i\bm{\phi}_i\quad\text{and}\quad p_h = \sum_iP_i\psi_i.
\end{displaymath}
The matrices have entries
\begin{align}
\label{eqn_fem_matrices}
  M_{ij} &= \langle\bm{\phi}_i,\bm{\phi}_j\rangle,& L_{ij} &=
a(\bm{\phi}_i,\bm{\phi}_j),\\
  G_{ij} &= -b(\bm{\phi}_j,\psi_i), & D_{ij} &= b(\bm{\phi}_i,\psi_j).
\end{align}
The forcing term $F$ can be decomposed into a contribution from the microstructure coupling and thermal fluctuations $F = F^{\mbox{\tiny str}} + F^{\mbox{\tiny thm}}$. The microstructure coupling term $F^{\mbox{\tiny str}}$ is given by $F_i^{\mbox{\tiny str}} = \langle\Fentrain,\bm{\phi}_i\rangle = \langle \Lambda\left\lbrack-\nabla \Phi \right\rbrack ,\bm{\phi}_i\rangle$ which by linearity of $\Lambda$ we can express as
\begin{align*}
F_i^{\mbox{\tiny str}} &= B_{\bX}^T(-\nabla\Phi).
\end{align*}
In the weak formulation, the operator $B_{\bX}^T$ now plays the analogous role as $\Lambda$ in equation~\ref{equ_f_int}.  
 
In summary, for the full fluid-structure system, the finite element semi-discretization can be expressed as
\begin{equation}  
  \begin{aligned}
    \rho M\dot{\mathbf{U}} &= -\mu L\mathbf{U} + GP + B_{\bX}^T(-\nabla\Phi) + \mb{F}^{\mbox{\tiny thm}}\\
  D\mathbf{U} &= 0\\
    \dot{\mathbf{X}} &= B_{\bX}\mathbf{U} \\
\label{equ_discr_SELM_full}
   \bar{\mb{F}}^{\mbox{\tiny thm}}  & = \langle \mb{F}^{\mbox{\tiny thm}} \rangle = k_{B}{T}\nabla_{\mathbf{X}} \cdot B_{\bX}^T  \\
   \left\langle \tilde{\mb{F}}^{\mbox{\tiny thm}}\left(\tilde{\mb{F}}^{\mbox{\tiny thm}}\right)^T\right\rangle &= \covG = \mu \left(LC + (LC)^T\right)  \hspace{3cm}  \\
   C &= k_B{T}\rho^{-1}M^{-1}. 
  \end{aligned}
\end{equation}
As in the continuum formulation, the thermal fluctuations are decomposed into the mean and fluctuating parts $\mb{F}^{\mbox{\tiny thm}} = \bar{\mb{F}}^{\mbox{\tiny thm}} + \tilde{\mb{F}}^{\mbox{\tiny thm}}$.
The operator $B_{\bX}$ now plays the analogous role as $\Gamma$ 
and $B_{\bX}^T$ the analogous role as $\Lambda$ in equations~\ref{eq:TDStokes}--~\ref{equ_f_int}.
This ensures that the important adjoint condition between the 
force-spreading and velocity-averaging operators hold~\cite{Peskin2002, AtzbergerSELM2011, AtzbergerSIB2007}.
We caution that the stochastic term $\mb{F}^{\mbox{\tiny thm}}$ was not derived by simply
projecting the full stochastic field $\mb{f}_{\mbox{\tiny thm}}$. Following the approach in~\cite{AtzbergerSELM2011,AtzbergerSIB2007}, we took into account the particular properties of the discrete operators of the system to ensure fluctuations propagate throughout the discretized fluid-structure system
in a manner consistent with statistical mechanics.   We establish in more detail the statistical mechanics of our finite element discretization in Section~\ref{sec_statMech}.

For the overdamped regime, we follow an approach similar to~\cite{AtzbergerTabak2013, AtzbergerSELM2011} to reduce the semi-discretized  equations~\ref{equ_discr_SELM_full} in the limit of rapid hydrodynamic relaxation to obtain a stochastic semi-discretization of the overdamped equations~\ref{equ_SELM_stokes}.  This yields
\begin{eqnarray}
\label{equ_overdamped_SELM}
\dot{\mathbf{X}} & = &  H(-\nabla\Phi) + k_B{T} \nabla_{\bX} \cdot H + \mathbf{R} \\
\langle \mathbf{R}\mathbf{R}^T \rangle & = & 2 k_B{T} H.
\end{eqnarray}
The effective hydrodynamic coupling tensor is $H = B_{\bX} S B_{\bX}^T$.  The operator $S$ represents solving for the fluid velocity $U$ in the following discretized Stokes equations
\begin{eqnarray}
\label{equ_Stokes_discr}
  \A
  \begin{bmatrix}
    \mathbf{U}\\P
  \end{bmatrix}
  =
  \begin{bmatrix}
    B_{\bX}^T(-\nabla\Phi)\\0
  \end{bmatrix},
  \quad\text{where}\quad
  \A =
  \begin{bmatrix}
    \mu L & -G\\
    D & 0
  \end{bmatrix}.
\end{eqnarray}
This can be expressed more explicitly as $S = [ I \hspace{0.1cm} 0 ] \A^{-1} \lbrack I \hspace{0.1cm} 0 \rbrack^T$
which gives $\mathbf{U} = SB_{\bX}^T(-\nabla\Phi)$.   

For general finite elements a semi-discretization of the stochastic fluid-structure system in the regimes of inertial dynamics and overdamped dynamics is given by 
equations~\ref{equ_discr_SELM_full} and~\ref{equ_overdamped_SELM}.  For use in practice, this semi-discretization requires a specific choice of appropriate approximating function spaces and corresponding finite elements.

\subsection{Choice of Finite Elements and the Approximating Function Spaces $\mathcal{V}_h$ and $\mathcal{P}_h$}
\label{sec_element_choice}
The Stokes problem given by equation~\ref{equ_Stokes_discr} for our mixed finite element discretization is well-known to be a saddle problem.   
To ensure numerical stability and a well-posed variational problem requires a careful choice for the two function spaces $\mathcal{V}_h$ and $\mathcal{P}_h$~\cite{Brenner2008}.  The Stokes problem can be split into two sub-problems.  In the first, one attempts to find the fluid velocity $\mathbf{u}_h$ within the subspace $\mathcal{Z} \subset \mathcal{V}_h$ of vector fields satisfying exactly the incompressibility constraint, $\mathcal{Z} = \{\mb{v} \in \mathcal{V}_h | b(\mb{v},q) = 0, \forall q\in\mathcal{P}_h\}$.  This is done by 
considering the variational problem~\cite{Brenner2008} 
\begin{eqnarray}
\label{equ_subpr_1}
a(\mb{u}_h,\boldsymbol{\phi}) & = & -\langle\bm{f},\bm{\phi}\rangle, \mbox{\hspace{0.4cm}} \forall \boldsymbol{\phi} \in \mathcal{Z} \subset \mathcal{V}_h.
\end{eqnarray}
In the second, one attempts to solve for the pressure $p$ in $\mathcal{P}_h$ by considering the variation problem~\cite{Brenner2008}  
\begin{eqnarray}
\label{equ_subpr_2}
b(\bm{\phi},p) & = & -a(\bm{u}_h,\bm{\phi}) - \langle\bm{f},\bm{\phi}\rangle, \mbox{\hspace{0.4cm}}  \forall \bm{\phi} \in \mathcal{V}_h.
\end{eqnarray}
Provided both of these two sub-problems can be solved, we have a consistent solution to the Stokes problem.
A central issue is that for mixed problems the bilinear form $b$ is often not
coercive.  This prevents a direct application of the Lax-Milgram Theorem to ensure well-posedness~\cite{Brenner2008}.
As an alternative, Babu{\v{s}}ka and Brezzi~\cite{Babuska1973,Hartmann1986,Brezzi1974} found  
for the two bilinear forms $a$, $b$ on the function spaces $\mathcal{V}_h$ and $\mathcal{P}_h$
a set of conditions that are weaker than coercitivity but still sufficient to ensure well-posedness.  
The first condition concerns the bilinear form $a$ and amounts to a form of coercitivity but now weaker only 
requiring this property when restricting to the subspace $\mathcal{Z}$ as suggested by the sub-problem 
in equation~\ref{equ_subpr_1}
\begin{eqnarray}
\alpha \|\mb{v} \|^2 & \leq & a(\mb{v},\mb{v}), \mbox{\hspace{0.4cm}}  \forall \mb{v} \in \mathcal{Z} \subset \mathcal{V}_h.
\end{eqnarray}
The second condition requires the bilinear form $b$ satisfy for the two function spaces
\begin{eqnarray}  
\label{equ_inf_sup}
\inf_{q_h\in\mathcal{P}_h}\sup_{\bm{v}_h\in\mathcal{V}_h}\frac{b(\bm{v}_h,q_h)}{
\|\bm{v}_h\|_{\mathcal{V}_h}  \|q_h\|_{\mathcal{P}_h}}  \geq\beta > 0.
\end{eqnarray}
This is referred to as the \textit{Babu{\v{s}}ka-Brezzi condition} or the \textit{inf-sup condition}.
Babu{\v{s}}ka and Brezzi showed for mixed methods these conditions are sufficient to ensure 
well-posedness of the variational problems and desirable numerical properties~\cite{Babuska1973,Hartmann1986,Brezzi1974}.

The Babu{\v{s}}ka-Brezzi condition provides important guidance on which finite elements should 
be used for the fluid and pressure in the Stokes problem.
For the Stokes problem, many of the most obvious choices of function spaces do not work very well in practice 
and in fact violate the Babu{\v{s}}ka-Brezzi condition given by 
equation~\ref{equ_inf_sup}~\cite{Arnold1984,Brenner2008}.  For instance using
$\mathcal{V}_h=(\mathbb{P}_k)^3$, $\mathcal{P}_h=\mathbb{P}_k$ where $\mathbb{P}_k$ is the space of
piecewise-polynomials of degree $k$ does not satisfy the Babu{\v{s}}ka-Brezzi condition for any choice of $k > 1$.
Even the choice $\mathcal{V}_h=(\mathbb{P}_1)^3$,$\mathcal{P}_h=\mathbb{P}_0$ does not satisfy the conditions 
to provide a stable method.

\begin{figure}[h]
\centering
\includegraphics[width=0.9\textwidth]{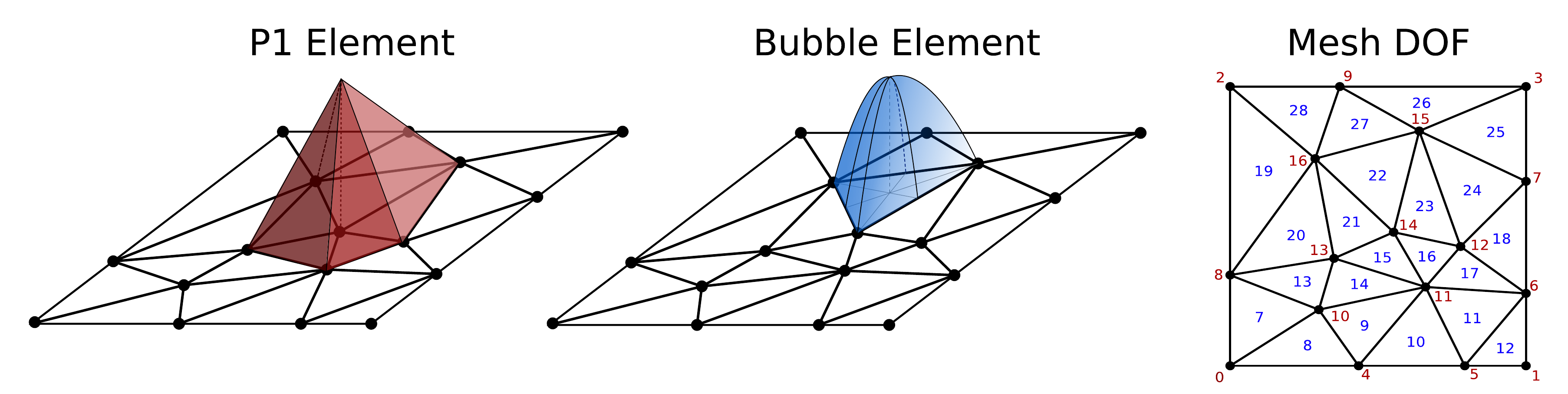} 
\caption{$\mathbb{P}_1$-MINI Elements~\cite{Arnold1984}.  To obtain a discretization of the Stokes equations satisfying 
the Babu{\v{s}}ka-Brezzi condition we use a combination of piecewise 
linear elements (P1-elements) for the pressure and quartic Bubble-elements for the velocity.  
The mesh degrees of freedom (DOF) consist of the usual nodal variables for P1-elements (labeled in red) 
and nodal variables at the center of each element to determine the bubble mode (labeled in blue).
While the bubble-elements contribute a large number of degrees of freedom, they have the convenient property
that their support is only within the interior of an element and are decoupled from one another~\cite{Arnold1984}.}  
\label{fig_FEM_Schematic}
\end{figure}

To satisfy the Babu{\v{s}}ka-Brezzi conditions, we use finite elements for the velocity field that are enriched by 
an additional bubble mode $\mathcal{V}_h=\left(\mathbb{P}_1\text{-\mbox{bubble}}\right)^3$ and 
for the pressure $\mathcal{P}_h=\mathbb{P}_1$, 
as introduced in~\cite{Arnold1984}.  
The $\mathbb{P}_1\text{-bubble}$ 
consists of the usual piecewise linear functions but enriched with a quartic ``bubble
element'' located at the barycenter of each tetrahedral element but vanishing at the faces, see 
Figure~\ref{fig_FEM_Schematic}.  
The enrichment by bubble-elements provides enough stabilisation that the Babu{\v{s}}ka-Brezzi conditions
are satisfied and yield for the Stokes equations a convergent method which has order $h$ \cite{Brenner2008}. 
Given that the bubble-elements do not overlap since they are contained within the interoir of each element,
the overhead associated with the bubble enrichment is marginal since they contribute in a manner 
to the overall linear system that is decoupled.  This also has desirable properties in the stochastic setting 
when generating fluctuations as we shall discuss.  The combination 
$\left(\mathbb{P}_1\text{-\mbox{bubble}}\right)^3$/$\mathbb{P}_1$ has a minimal footprint.  
We shall refer throughout to this pair as the $\mathbb{P}_1$-MINI Elements, as in~\cite{Arnold1984}.

\subsection{Statistical Mechanics of the Semi-Discretization}
\label{sec_statMech}
We show that our stochastic semi-discretization provides an approximation yielding fluctuations consistent with statistical mechanics.  We show that the Gibbs-Boltzmann distribution is invariant under the stochastic dynamics.  The Gibbs-Boltzmann distribution is given by
\begin{eqnarray}
\label{equ_GB_distr}
\Psi_\subtxt{GB}(\mathbf{z}) = \frac{1}{Z}\exp\left[{-E(\mathbf{z})/k_B{T}}\right].
\end{eqnarray}
The $\mathbf{z} = (\mathbf{u},\mathbf{X})$ is the state of the system,
$E$ is the energy, $k_B$ is Boltzmann's constant, $T$
is the system temperature, and $Z$ is a normalization constant for the distribution~\cite{Reichl1998}.  The energy of the fluid-structure system is 
given by the kinetic energy of the fluid and the potential energy of the microstructures 
\begin{eqnarray}
E[\mathbf{z}] & = & \frac{\rho}{2}  \int |\mathbf{u}|^2 dx + \Phi(\mathbf{X}).
\end{eqnarray}
For the discrete system we consider the energy
\begin{eqnarray}
\label{equ_discr_energy}
E[\mathbf{z}] & = & \frac{\rho}{2} \mathbf{U}^T M \mathbf{U} + \Phi(\mathbf{X}).
\end{eqnarray}
The $M$ is the mass matrix defined in equation~\ref{eqn_fem_matrices}.  For this energy, we have at equilibrium that the fluid velocity 
$\mathbf{U}$ has fluctuations with mean zero and covariance 
\begin{eqnarray}
\label{equ_cov_U}
C = k_B{T}\rho^{-1}M^{-1}.
\end{eqnarray}
For the stochastic fluid-structure equations~\ref{equ_discr_SELM_full}, the Fokker-Planck equation for $\Psi_\subtxt{GB}$ is given by
\begin{eqnarray}
\label{equ_flux_II}
\frac{\partial \Psi_\subtxt{GB}}{\partial t} & = & - \nabla \cdot \mathbf{J} \\
\nonumber
\mathbf{J} & = &
\left[ 
\begin{array}{l}
\rho^{-1} M^{-1} \left( \mu L + B_{\mb{X}}^T + k_B{T} \nabla_{\mb{X}}\cdot B_{\mb{X}}^T \right) \\
B_{\mb{X}}
\end{array}
\right]\Psi_\subtxt{GB} - \frac{1}{2}(\nabla \cdot \covG)\Psi_\subtxt{GB} - \frac{1}{2} \covG \nabla \Psi_\subtxt{GB}.
\end{eqnarray}
The $\covG$ denotes the covariance operator for the stochastic driving fields given by equation~\ref{equ_discr_SELM_full}.
The Gibbs-Boltzmann distribution is invariant provided that 
\begin{eqnarray}
\label{equ_appendix_div_J2}
\nabla \cdot \mathbf{J} & = & A_1 + A_2 + \nabla\cdot \mathbf{A}_3 + \nabla\cdot \mathbf{A}_4 = 0
\end{eqnarray}
where
\begin{eqnarray}
\nonumber
A_1            & = & \left[\rho^{-1} M^{-1} \left(B_{\mb{X}}^T + k_B{T} \nabla_{\mb{X}}\cdot B_{\mb{X}}^T \right)\cdot\nabla_{\mathbf{U}}E 
+ B_{\mb{X}}\cdot\nabla_{\mathbf{X}}E
\right](-k_B{T})^{-1}\Psi_\subtxt{GB} \\
\nonumber
A_2            & = & \left[\nabla_{\mathbf{U}}\cdot\left(\rho^{-1} M^{-1} \left(B_{\mb{X}}^T + k_B{T} \nabla_{\mb{X}}\cdot B_{\mb{X}}^T \right)\right) 
+ \nabla_{\mathbf{X}}\cdot B_{\mb{X}}\right]\Psi_\subtxt{GB} \\
\nonumber
\mathbf{A}_3   & = & -\frac{1}{2}(\nabla \cdot \covG)\Psi_\subtxt{GB} \\
\nonumber
\mathbf{A}_4   & = & \left[
\begin{array}{l}
\mu L +\left[{\covG_{\mb{U}\mb{U}}\nabla_{\mathbf{U}}E +
\covG_{\mb{U}\mb{X}}\nabla_{\mathbf{X}}E}\right]({2k_B{T}})^{-1}
\\
\left[
\covG_{\mb{X}\mb{U}}\nabla_{\mathbf{U}}E + \covG_{\mb{X}\mb{X}}\nabla_{\mathbf{X}}E\right]({2k_B{T}})^{-1} \\
\end{array}
\right]\Psi_\subtxt{GB}.
\end{eqnarray}
To simplify the notation we have suppressed explicitly denoting the functions on which the 
operators act, which can be inferred from equation~\ref{equ_discr_SELM_full}.  We have also suppressed the incompressibility constraint since this can be handled readily by considering 
the dynamics decomposed into a component on the space of solenoidal fields and its orthogonal complement as was done in~\cite{AtzbergerTabak2013}.
To compare terms, we use the gradients obtained from the energy in equation~\ref{equ_discr_energy} given by
\begin{eqnarray}
\label{equ_appendix_grad_p_E2}
\nabla_{\mb{U}}E & = & \rho M \mathbf{U}\\
\label{equ_appendix_grad_X_E2}
\nabla_{\mb{X}}E & = & \nabla_{\mb{X}}\Eng.
\end{eqnarray}
By direct substitution of the gradients into $A_1$ given by 
equation~\ref{equ_appendix_grad_p_E2}--~\ref{equ_appendix_grad_X_E2},
we find after cancellation 
$A_1 
= -(\rho^{-1}M^{-1}(\nabla_{\mb{X}} \cdot B_{\mb{X}}^T)\cdot\nabla_{\mb{U}}E) \Psi_\subtxt{GB} 
= -(\mb{U}^T(\nabla_{\mb{X}} \cdot B_{\mb{X}}^T)) \Psi_\subtxt{GB} $.
Since $B_{\mb{X}}^T$ only depends on $\mb{X}$, we have 
$A_2 = (\nabla_{\mb{X}}\cdot B_{\mb{X}}\mb{U})\Psi_\subtxt{GB}$.
This gives that $A_1 + A_2 = 0$.  We remark that this is a direct 
consequence of requiring the coupling operators to be linear and 
adjoints $\OpFSCouple = \OpSFCouple^{\OpAdjoint}$ in the sense of equation~\ref{equ_adjoint} where 
$\OpFSCouple = B_\mb{X}$~\cite{AtzbergerSELM2011,AtzbergerTabak2013}. 
 The term $\mb{A}_3$ accounts for 
probability fluxes driven by state dependent changes in the covariance of the 
stochastic driving fields.  In this case the 
covariance $\covG$ does not depend on the system state, so 
the divergence gives $\mb{A}_3 = 0$.  
The term $\mb{A}_4$ arises from the interplay between 
dissipation and fluctuations in the dynamics.  By looking at the 
dependence of the differentiated expressions most of
the terms are seen to be zero.   By our choice of the covariance 
$\covG_{\mb{u},\mb{u}}$ given by 
equation~\ref{equ_discr_SELM_full} which was 
motivated by the form of the discrete energy 
in equation~\ref{equ_discr_energy} and the target equilibrium 
covariance for fluctuations given in equation~\ref{equ_cov_U}, 
we have that the non-zero terms balance to yield $\mb{A}_4 = 0$.
The choice of $\covG$ and its relation to $C$ can be thought of as 
imposing a discrete fluctuation-dissipation balance for our 
semi-discretization~\cite{AtzbergerSIB2007, AtzbergerSELM2011}.
These results establish that $\nabla\cdot\mb{J} = 0$ and 
that the Gibbs-Boltzmann distribution is invariant under
our semi-discretized stochastic dynamics.

In the overdamped limit, the fluid-structure system is governed only by the microstructure potential energy 
\begin{eqnarray}
\label{equ_discr_energy2}
E[\mathbf{X}] & = & \Phi(\mathbf{X}).
\end{eqnarray}
For the stochastic fluid-structure equations~\ref{equ_overdamped_SELM},
the Fokker-Planck equations  for $\Psi_\subtxt{GB}$ are given by 
\begin{eqnarray}
\label{equ_flux_overdamped}
\frac{\partial \Psi_\subtxt{GB}}{\partial t} & = & - \nabla \cdot \mathbf{J} \\
\nonumber
\mathbf{J} & = & \left(H (-\nabla_{\mb{X}} \Phi(\mb{X})) + k_B{T} \nabla \cdot H \right)\Psi_\subtxt{GB} 
- \frac{1}{2}(\nabla \cdot 2k_B {T} H)\Psi_\subtxt{GB} - \frac{1}{2} (2k_B{T}H) \nabla_{X} \Psi_\subtxt{GB}.
\end{eqnarray}
From the form of $\Psi_\subtxt{GB}$, we have 
$-\frac{1}{2} (2k_B{T}H) \nabla_{X} \Psi_\subtxt{GB} = H (\nabla_{\mb{X}} \Phi(\mb{X})) \Psi_\subtxt{GB}$ which gives
that $\mathbf{J} = 0$.  This shows that the Gibbs-Boltzmann distribution is invariant with detailed-balance
under the stochastic dynamics of our semi-discretization given in equation~\ref{equ_overdamped_SELM}.

These results establish that our semi-discretizations both in the inertial and overdamped regimes yield stochastic dynamics consistent with statistical mechanics.
An important issue in practice is to generate efficiently the stochastic driving fields $\mathbf{F}^{thm}$ and $\mathbf{h}$ with the 
required covariances.

\section{Generation of the Stochastic Driving Fields}
\label{sec_stoch_field_gen}
The thermal fluctuations in the overdamped regime require generating a  
Gaussian random field $R$ with mean zero and covariance 
\begin{eqnarray}
\langle \mb{R} \mb{R}^T \rangle = 2k_BT H = 2k_BTB_{\bX}SB_{\bX}^T. 
\end{eqnarray}
In general, this is difficult since $S$ is a dense matrix 
and calculating the square root using methods such as Cholesky yields a dense factor 
making field generation computationally expensive.  The issue is further complicated 
by the fact that the action of $S$ involves taking the inverse of $\A$ which is not a 
sign definite matrix.  We take an alternative approach for the efficient generation of the stochastic
field $\mb{R}$ by factoring the Stokes operator $S$ to reduce the problem to generating variates with
a covariance related only to the discrete Laplacian stiffness matrix $L$.

\subsection{Splitting the Covariance using Properties of the Stokes Operator and Laplacian}
A useful identity for the discrete Stokes operator is
\begin{eqnarray}
S(\mu L)S = S. 
\end{eqnarray}
This follows since for an arbitrary choice of $F^1$ in equation~\ref{equ_overdamped_SELM} we have the solution 
$U_0 = SF^1$ so that $S(\mu L)SF^1 = SF^1 + S G P_0$ from substitution of $U_0$ back 
into the Stokes equation~\ref{equ_overdamped_SELM}.  The $S G = 0$ since for an arbitrary 
$P_1$ if we set $F^{1} = G P_1$ then the solution is 
clearly $U_1 = S F^{1} = S G P_1 = 0$ by uniqueness the Stokes problem for $U$.

This identity is useful since it reduces generating the stochastic driving field $\mathbf{R}$ with 
covariance $2k_B{T}H$ to that of generating variates $\bm{\xi}$ with covariance $C = -(\mu L)$.  
More specifically, we can generate the stochastic driving field by using $\mathbf{R} = \sqrt{2k_BT}B_{\bX}S\bm{\xi}$ 
which yields the required statistics through
\begin{align*}
  \langle \mathbf{R} \mathbf{R}^T\rangle &= 2k_BTB_{\bX}S\langle \bm{\xi}\bm{\xi}^T\rangle S^TB_{\bX}^T
  = 2k_BTB_{\bX}S(\mu L)SB_{\bX}^T
  = 2k_BTB_{\bX}SB_{\bX}^T = 2k_BT H.
\end{align*}
To generate efficiently the variates $\bm{\xi}$ with covariance $-L$, we develop stochastic iterative methods.

We remark that for this factorization there are some further advantages when using the $\mathbb{P}_1$-MINI elements.  
For the Laplacian stiffness matrix $L$, the collection of entries corresponding 
to the bubble elements is diagonal.  This allows for trivial generation of variates 
at the bubble elements by using independent Gaussians and weighting by the square root of the diagonal entry of the stiffness matrix.  
This means that stochastic field generation can be reduced to the problem of generating variates with covariance $-L$ restricted 
to the non-bubble nodal degrees of freedom which are far fewer 
in number than the bubble nodes, see Figure~\ref{fig_FEM_Schematic}.  This provides in the stochastic setting a 
particular advantage of the $\mathbb{P}_1$-MINI elements over higher order methods such as the 
Taylor-Hood elements~\cite{Arnold1984,Brenner2008}.  

\subsection{Stochastic Iterative Methods}
We develop stochastic iterative methods to obtain Gaussian random variates with a target mean and covariance structure.
A general one-step linear iterative method can be developed into a Gibbs sampler by introducing 
a stochastic term $\bm{\eta}^n$ into the iteration step as
\begin{eqnarray}
\label{equ_stoch_iter}
\bm{\xi}^{n + 1}  = M\bm{\xi}^{n} + N\mathbf{b} + \bm{\eta}^n.
\end{eqnarray}
We assume throughout that $\bm{\eta}^n$ is a Gaussian variate with mean zero and covariance $G = \langle \bm{\eta}^n \bm{\eta}^{n,T} \rangle$.
The stochastic iterations can be expressed in terms of the probability density $\rho^{n}(\bm{\xi})$ and transitions at iteration $n$ as
\begin{eqnarray}
\rho^{n + 1}(\bm{\xi}) & = & \int \pi(\bm{\xi},\mathbf{w}) \rho^{n}(\mathbf{w}) d\mathbf{w} \\
\pi(\bm{\xi},\mathbf{w}) & = & \frac{1}{\sqrt{2\pi \mbox{det}G}}
\exp\left[ \left(\bm{\xi} - M\mathbf{w} - N\mathbf{b}\right)^TG^{-1}\left(\bm{\xi} - M\mathbf{w}  - N\mathbf{b}\right) \right].
\end{eqnarray}

This yields a set of Gaussians $\bm{\xi}^n$ having mean value $\bm{\mu}^n$ and covariance $C^n$ satisfying
\begin{eqnarray}
\bm{\mu}^{n + 1} & = &  M\bm{\mu}^{n} + N\mathbf{b} \\
C^{n + 1}   & = &  MC^{n}M^T + N\mathbf{b}\bm{\mu}^{n,T}M^T + M \bm{\mu}^n \mathbf{b}^T N^T + N\mathbf{b}\mathbf{b}^TN^T + G.
\end{eqnarray}

In the case that the target mean $\bm{\mu} = 0$, we can choose $\mathbf{b} = 0$ and this simplifies to
\begin{eqnarray}
\label{equ_Cnp1_simple}
C^{n + 1} & = &  MC^{n}M^T + G.
\end{eqnarray}
The autocorrelation $\Phi^{m} = \langle \bm{\xi}^n (\bm{\xi}^{n + m})^T\rangle$ for how the random variates $\bm{\xi}^n$ decorrelate over $m$
iterations satisfies
\begin{eqnarray}
\label{equ_iter_autoCorr}
\Phi^{m + 1} & = & M\Phi^{m}.
\end{eqnarray}
This relation has the consequence that 
in the stochastic setting the decay in correlation has the same behavior as the decay in error in the deterministic setting when iteratively solving the problem $L\mathbf{x} = \mathbf{b}$. 
To obtain random variates with a target covariance $C$, we have from equation~\ref{equ_Cnp1_simple} that the covariance of $\bm{\eta}$ must satisfy 
\begin{eqnarray}
\label{equ_iter_G_C_relate}
G = C - MCM^T = -AC - CA^T +ACA^T
\end{eqnarray}
where $A$ is defined by $M = I - A$.  
For generating general Gaussian random variates $\bm{\xi}$, the stochastic iterative method~\ref{equ_stoch_iter} provides a useful approach provided 
the iterative method converges efficiently and 
the random variates $\bm{\eta}$ have a covariance $G$ that is easier to generate than $C$.  For 
traditional iterative methods, such as SOR, Gauss-Seidel, and Jacobi iterations, stochastic iterative counterparts have been introduced in~\cite{Goodman1989,Goodman1986,Adler1981,Whitmer1984}.  The convergence of such stochastic iterative methods can be further improved by using preconditioning strategies such as multigrid~\cite{Goodman1989,Goodman1986}.  For stochastic field generation in our semi-discretized fluctuating hydrodynamic equations, we show how the multigrid preconditioner can be adopted to sample the stochastic driving fields.  We base our sampler on
Gauss-Seidel iterations.

\subsection{Modified Gauss-Seidel Iterations for Gibbs Sampling}
The Gauss-Seidel iterative method for solving the system $A\mb{v} = \mb{b}$ in the deterministic setting is given by 
\begin{eqnarray}
\mb{v}^{n + 1}  = (D - L)^{-1}U\mb{v}^{n} + (D - L)^{-1}\mathbf{b}.
\end{eqnarray}
We use the splitting $A = D - L - U$.  The $D$ denotes the diagonal elements of the matrix $A$, $L$ the lower triangular elements, and $U$ the upper triangular elements.
This corresponds to the general iterative method of equation~\ref{equ_stoch_iter} with 
the choice $M = (D - L)^{-1}U$ and $N = (D - L)^{-1}$.
To obtain Gaussian random variates with mean $\bm{\mu} = 0$ and covariance $C = A^{-1}$, we must generate for each iteration a stochastic 
variate $\bm{\eta}^n$ having covariance $G$ satisfying equation~\ref{equ_iter_G_C_relate}.  For Gauss-Seidel iterations, this requires covariance
\begin{eqnarray}
\label{equ_G_cov_GS}
G = (D - L - U)^{-1} - (D - L)^{-1}U (D - L - U)^{-1} L (D - L)^{-T}.  
\end{eqnarray}
It has been shown in~\cite{Goodman1989} that this can be expressed as
\begin{eqnarray}
\nonumber
G  
     & = & (D - L)^{-1} \left[ (D - L - U)(D - L - U)^{-1} (D - U)   \right.\\
      && \hspace{2.45cm} \left.+ \hspace{0.125cm} U (D - L - U)^{-1} (D - L - U)
     \right] (D - U)^{-1} \\     
     \nonumber
     & = & (D - L)^{-1} D (D - U)^{-1}.
\end{eqnarray}
This gives the factorization $G = QQ^T$ with 
\begin{eqnarray}
\label{equ_def_Q_GS}
Q & = & (D - L)^{-1} D^{1/2}.
\end{eqnarray}
We have used that $A$ is symmetric positive definite with $U = L^T$ and that the transpose of the inverse is the 
inverse of the transpose.  This provides a way to generate the random variates through $\bm{\eta} = Q\bm{\zeta}$ where
$\bm{\zeta}$ is a standard multi-variate Gaussian with mean zero and unit covariance.  This is an efficient procedure, especially for sparse matrices $A$, since the action of $(D - L)^{-1}$ can be obtained through back-substitution in the same manner as in deterministic Gauss-Seidel iterations~\cite{Briggs2000}.  This provides our stochastic Gauss-Seidel iterative method for the generation of random variates.  

\subsection{Stochastic Multigrid}
\begin{figure}[h]
\centering
\includegraphics[width=0.9\textwidth]{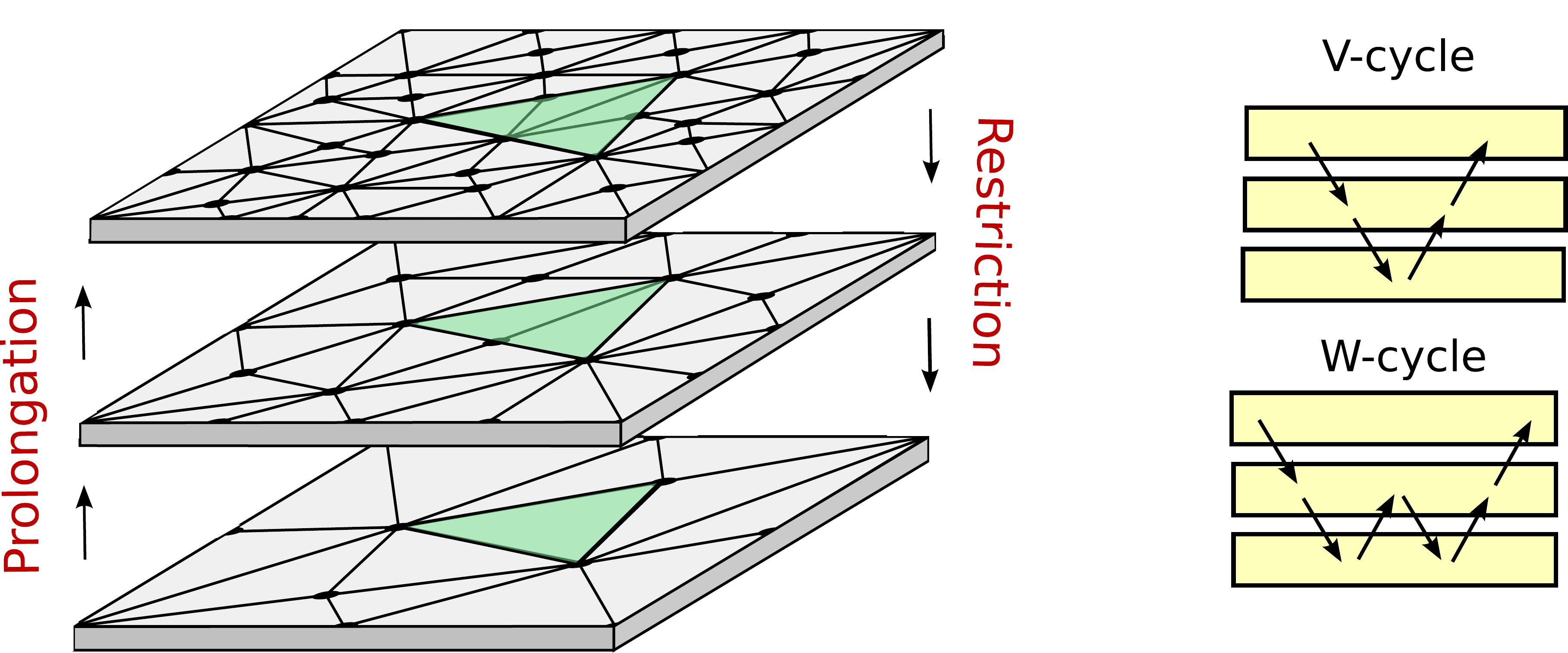}
\caption{Multigrid uses subproblems on different levels of refinement to iteratively solve the system 
of equations.  Three operators are utilized: (i) \textit{smoother operator}, (ii) \textit{restriction operator}, 
and (iii) \textit{prolongation operator}.  The \textit{smoother operator} serves on each level to iteratively relax values toward the solution.  The \textit{restriction operator} and \textit{prolongation operator} serve to transfer data between levels (left).  Multigrid iterations are performed 
by a protocol combining these operations (right)~\cite{Briggs2000}.
}
\label{fig_Stochastic_ML}
\end{figure}

In the deterministic setting multigrid is often used to improve the convergence of iterative methods~\cite{Briggs2000,McCormick1989}. 
Multigrid makes use of different levels of refinement $\ell$ to perform iterations.  In the multigrid iterations three fundamental operators are utilized. The first is the \textit{smoother operator} $\mathcal{S}^{\ell}$ which is used to iteratively approximate the linear system of equations on a given level $\ell$. For transmission of information between different levels of resolution, a \textit{restriction operator} $\mathcal{I}_{\ell}^{\ell - 1}$ and \textit{prolongation operator} $\mathcal{I}_{\ell - 1}^{\ell}$ are defined.  The restriction operator $\mathcal{I}_{\ell}^{\ell - 1}$ maps data from a more refined level $\ell$ to data on a less refined level $\ell - 1$.  The prolongation operator $\mathcal{I}_{\ell - 1}^{\ell}$ maps data from a less refined level $\ell - 1$ to data on a more refined level $\ell$,  see Figure~\ref{fig_Stochastic_ML}. 
Multigrid provides improvements in the convergence in the deterministic setting by the different rates that the smoother relaxes modes on different levels.  This same feature that relaxes errors translates to the stochastic setting by improving the rate at which the generated random variates decorrelate in the number of iterations.

To ensure a viable stochastic multigrid method it is useful to have a few additional properties that are not strictly required in the deterministic setting.  We take the linear equations $A^{(\ell)} \mathbf{v} = \mathbf{b}^{(\ell)}$ at refinement level $\ell$ to be related to those at the most refined level $\ell^*$ by 
\begin{eqnarray}
A^{(\ell)} = \mathcal{I}_{\ell^*}^{\ell} A\mathcal{I}_{\ell}^{\ell^*}, \hspace{1cm} \mathbf{b}^{(\ell)} = \mathcal{I}_{\ell^*}^{\ell} \mathbf{b}. 
\end{eqnarray}
An important property when adopting the multigrid method to the stochastic setting is to ensure a consistent variational principle for the linear equations at different refinement levels. The variational principle at the most refined level is that the solution is a minimizer of the energy
\begin{eqnarray}
\label{equ_energy_v}
E(\mathbf{v}) = \frac{1}{2}\mathbf{v}^T A \mathbf{v} 
               - \mathbf{v}^T\mathbf{b}.
\end{eqnarray}
Consistency requires that the energy defined by $E^{\ell}(\mathbf{v}^{(\ell)}) := E(\mathcal{I}_{\ell}^{\ell*}\mathbf{v}^{(\ell)})$
provide the variational principle at level $\ell$.  
This is satisfied when the prolongation and restriction operators are
adjoints $\mathcal{I}_{\ell^*}^{\ell} = \left(\mathcal{I}_{\ell}^{\ell^*}\right)^T$, which yields
\begin{eqnarray}
E^{\ell}(\mathbf{w}) &=& \frac{1}{2}\mathbf{w}^T \left(\mathcal{I}_{\ell}^{\ell^*}\right)^T A \left(\mathcal{I}_{\ell}^{\ell^*}\right)\mathbf{w} 
               - \mathbf{w}^T \left(\mathcal{I}_{\ell}^{\ell^*}\right)^T \mathbf{b} \\
& = &
\frac{1}{2} \mathbf{w}^T
A^{(\ell)}
\mathbf{w}
- \mathbf{w}^T \mathbf{b}^{(\ell)}. 
\end{eqnarray}
The variational property has some important consequences.  For the target Gaussian distribution $\rho(\mb{v})$, 
the stochastic smoother samples at level $\ell$ the marginal 
probability distribution $\rho_{(\ell)}(\mb{w}) = \int_{\{\mb{v} = \mathcal{I}_{\ell}^{\ell^*} \mb{w}\}} \rho(\mb{v}) d\mb{v}$, which is given by  
\begin{eqnarray}
\rho_{(\ell)}(\mathbf{w}) = 
\frac{1}{\sqrt{2\pi \mbox{det} A^{(\ell)} } }
\exp
\left[
-\frac{1}{2}
\mathbf{w}^T A^{(\ell)}\mathbf{w}
 + \mathbf{w}^T \mathbf{b}^{(\ell)} 
\right].
\end{eqnarray}
This variational property can be shown to be sufficient to ensure the probability distribution of the target multi-variate Gaussian is the invariant distribution of the stochastic multigrid iterations~\cite{Goodman1989}.

There are at least two common approaches taken when developing multigrid methods.  The first
is \textit{geometric multigrid} in which the restriction and prolongation operators are constructed
by considering geometric correspondences between different levels of the discretization such as 
spatial averages or interpolations~\cite{Briggs2000}.  The second is
\textit{algebraic multigrid} in which the restriction and prolongation operators are constructed by 
considering the algebriac structure of the linear system such as grouping clusters of entries of the matrix $A$ ~\cite{Elman2003504,Tuminaro00parallelsmoothed,Vanek98convergenceof,vanek1999two,vanvek1996algebraic}.  
We use an \textit{algebraic multigrid} method with \textit{smoothed aggregation} following an 
approach similar to~\cite{Elman2003504}.  

\newpage
\clearpage

\section{Algorithm : Summary of SELM based on FEM Stokes}
\label{sec_summary_alg}

\verb| |\\
\\
{\bf Input}: Polyhedral domain $\Omega$, initial body configuration $\bX_0$, body potential $\Phi$, terminal time $t_{\text{end}}$, timestep size $\Delta t$, Peskin $\delta$ footprint $h$, kinematic viscosity $\mu$, temperature $T$.\\
\\
{\bf Output}: Body configuration $\bX_t$ at each $t= \Delta t, 2\Delta t, 3\Delta t,\ldots$\\
\begin{enumerate}
\item Construct the adapted tetrahedralization: \\
  $\{\mathcal{T},V\}$ = \verb|AdaptedMesh|$(\Omega, \bX_0, h)$.
\item Construct the linear operators: $\{L,D\}$ = \verb|FluidOperators|$(\{\mathcal{T}, V\})$.
\item Set $\mathbb{A} = \begin{bmatrix}\mu L & D^T\\D & 0\end{bmatrix}$.
\item Set $t,t_{\text{remesh}}=0$.
\item Begin computing the body trajectory. While $t < t_{\text{end}}$, do steps (a)-(g):
  \begin{enumerate}
  \item Construct the velocity interpolator/force spreading operator:\\
    $B$ = \verb|VelocityInterpolator|$(\{\mathcal{T}, V\}, \bX_t)$.
  \item Generate the $N(\bm{0},(\mu L)^{-1})$ sample: $\bm{\xi}$ = \verb|GaussianSample|$(\mu L)$.
  \item Generate the body forces: $\bm{f} = B^T(-\nabla\Phi(\bX_t))$.
  \item Perform the fluid solve: $\bm{U}$ = \verb|FluidSolve|$(\mathbb{A}$, $\bm{f}\Delta t + \sqrt{2k_BT\Delta t}\mu L\bm{\xi})$
  \item Interpolate fluid velocity to particles: $\bX_{t+\Delta t} = \bX_t + B\bm{U}$
  \item Test for remeshing. If $|\bX^i_{t+\Delta t}-\bX^i_{t_{\text{remesh}}}| > 3h$ for any $i$, do steps i-iv:
    \begin{enumerate}
    \item Reconstruct the (adapted) tetrahedralization: \\
      $\{\mathcal{T},V\}$ = \verb|AdaptedMesh|$(\Omega, \bX_{t+\Delta t}, h)$.
    \item Reconstruct the linear operators: \\
      $\{L,D\}$ = \verb|FluidOperators|$(\{\mathcal{T}, V\})$.
    \item Set $\mathbb{A} = \begin{bmatrix}\mu L & D^T\\D & 0\end{bmatrix}$.
    \item Set $t_{\text{remesh}} = t+\Delta t$.
    \end{enumerate}
  \item Update the current time: set $t = t + \Delta t$.
  \end{enumerate}
\end{enumerate}

\newpage
\clearpage

\section{Validation Studies}
\label{sec_validation}
We perform several studies to establish the validity of the computational methods.
We first investigate the empirical covariance structure obtained from our iterative Gibbs samplers to make comparisons with the target covariance structure.  We then consider the decorrelation rates exhibited by the direct stochastic Gauss-Seidel iterations in comparison to the stochastic multigrid iterations.  We next explore the effective hydrodynamic coupling tensors corresponding to the case when 
the fluid and structures are coupled based on the \textit{Immersed Boundary Method}~\cite{Peskin2002,Bringley2008}.  We then investigate the equilibrium 
fluctuations of the fluid-structure system by computing the statistics of diffusing particles subject to a harmonic potential.  To validate 
our overall method we compare these results with the predicted Gibbs-Boltzmann distribution.

\subsection{Covariance Structure}
\begin{figure}[h]
\centering
\includegraphics[width=0.9\textwidth]{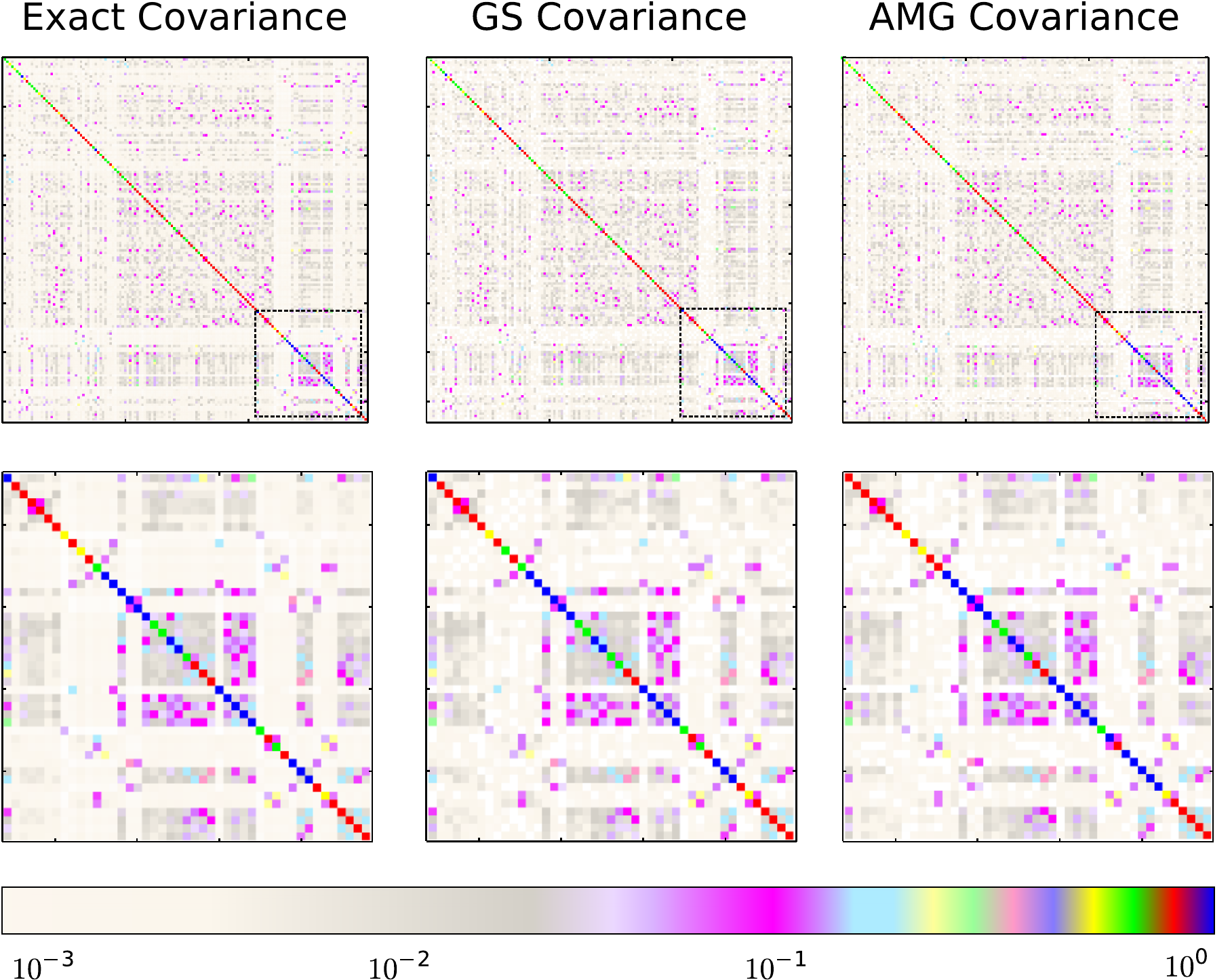}
\caption{Covariance Structure.  The matrix of covariance for all finite element degrees of freedom are shown on a logarithmic scale for 10,000 samples from our stochastic Gauss-Seidel iterations (GS) and stochastic multigrid iterations (AMG) (top row). These are compared to the target covariance $-L^{-1}$ which is the negative inverse Laplacian.  The diagonal entries show the self-correlation while the off-diagonal entries show correlations between distinct degrees of freedom.  The lower right entries of the correlation matrix is shown in more detail on the bottom.  Both stochastic iterative methods yield results with good agreement with the target covariance structure with an average error of 4\%.
}
\label{fig_CoVarStructure_matrix}
\end{figure}

A key component of our computational methods for fluctuating hydrodynamics is to generate efficiently the stochastic fields driving fields.  As discuss in Section~\ref{sec_stoch_field_gen}, this requires methods to generate Gaussian variates with target covariance $-L^{-1}$.  We consider the fluctuating hydrodynamics when confined within a spherical domain with Dirichlet no-slip boundary conditions.  For both direct stochastic Gauss-Seidel iterations and stochastic multigrid iterations we generated 10,000 samples, shown in Figure~\ref{fig_CoVarStructure_matrix}.  The covariance matrix is shown for all of the finite element degrees of freedom of the discrete system including the bubble modes.  The magnitude of the covariance is shown on a logarithmic scale.  We find that the empirical covariance structure from both of our stochastic iterative methods 
is in very good agreement with the target covariance structure with an average error of less than 4\%.  An important consideration is how efficiently the stochastic iterative methods can sample nearly independent Gaussian random variates.

\newpage
\clearpage

\subsection{Decorrelation Rates}
\begin{figure}[h]
\centering
\includegraphics[width=0.45\textwidth]{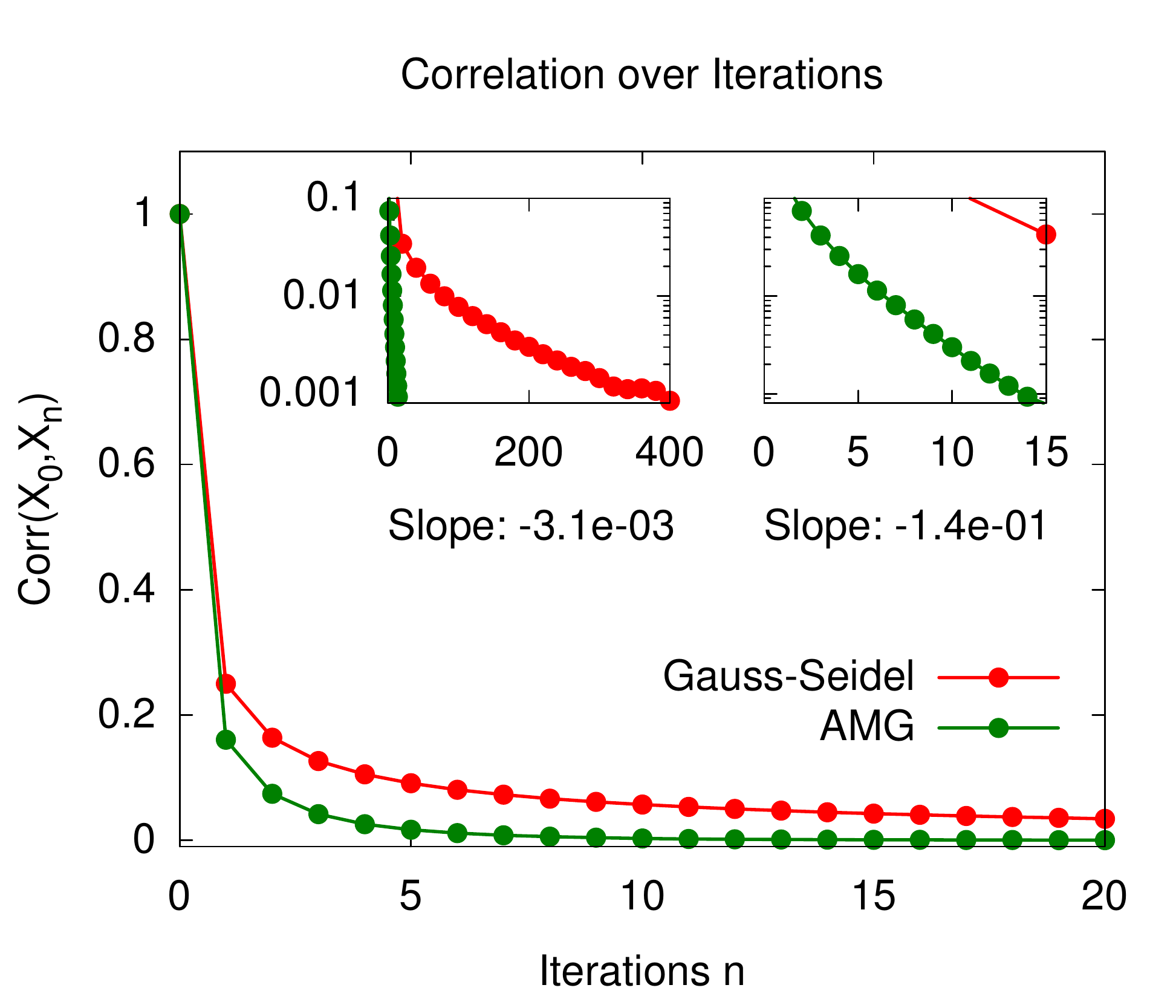}
\includegraphics[width=0.45\textwidth]{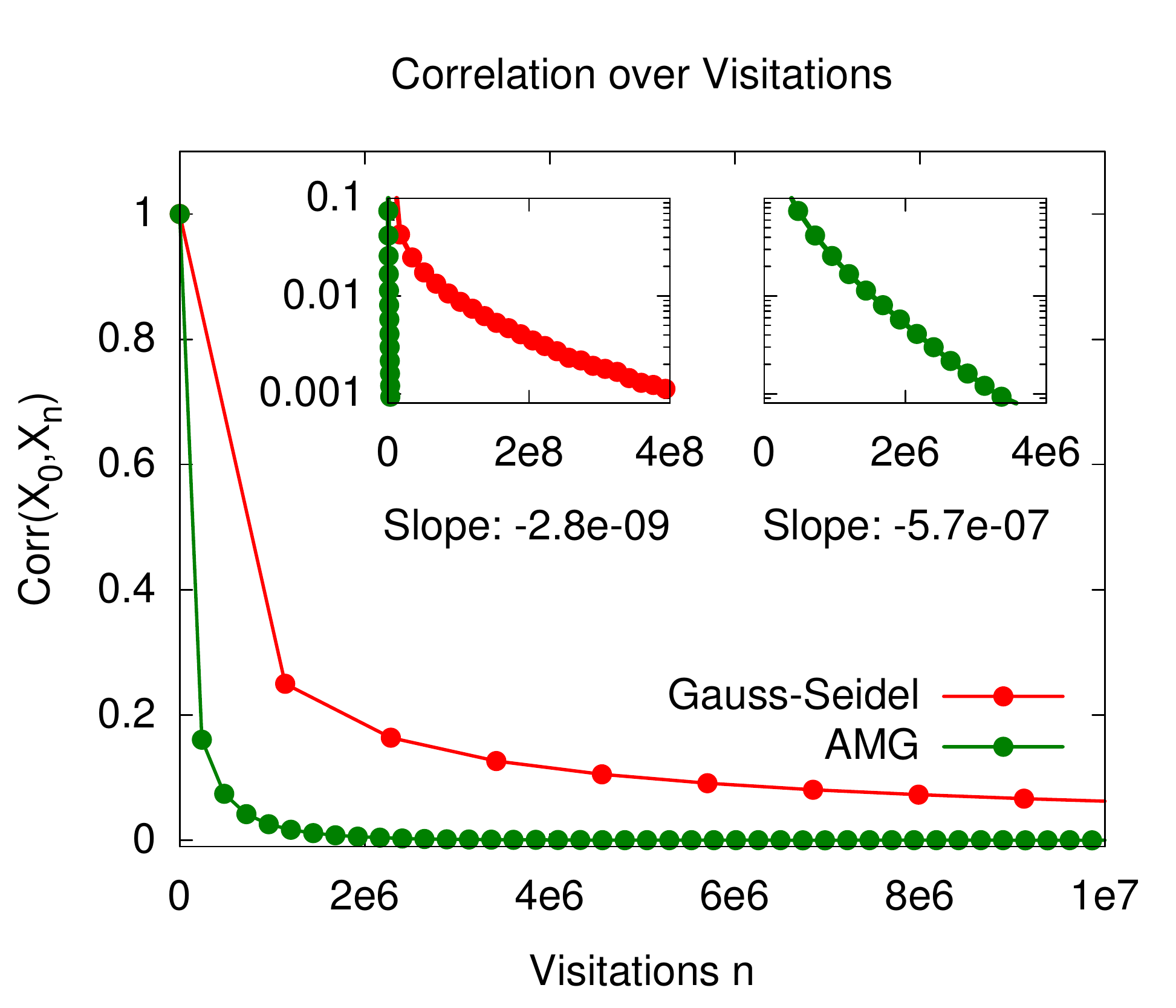}
\caption{Decorrelation of random variates.  Sampling was perform for the stochastic driving fields 
on a spherical domain with Dirichlet boundary conditions for the target covariance of the discrete inverse Laplacian $-L^{-1}$.  We used both 
the stochastic Gauss-Seidel and stochastic multigrid methods.  The decorrelation in the number of iterations is shown on the left.  The decorrelation in the number of Gauss-Seidel visitations to the individual nodes of the finite element are shown on the right.  We find that the stochastic multigrid method greatly improves performance.  To achieve a correlation of less than $1\%$ between random variates, direct stochastic Gauss-Seidel requires on the order of $100$ iterations.  In contrast, stochastic multigrid requires less than $10$ iterations.  For the number of visitations, the stochastic Gauss-Seidel requires on the order of $10^8$ steps.  In contrast, the stochastic multigrid iterations only require $10^6$ steps.  The rate of exponential decay for stochastic Gauss-Seidel is estimated to be $-3.1\times 10^{-3}$ and the rate for the stochastic multigrid iterations $-1.4\times 10^{-1}$.  }
\label{fig_decorr_times}
\end{figure}

The stochastic driving fields appearing in the fluctuating hydrodynamic equations~\ref{eq:TDStokes} and~\ref{equ_SELM_stokes} are treated as uncorrelated in time.  In stochastic numerical methods this requires on successive time steps the generation of independent Gaussian random variates to account for the thermal fluctuations.  When using stochastic iterative methods to sample random variates there are always correlations between generated variates but these diminish over successive iterations.  This means the quality of the random variates for fluctuating hydrodynamic simulations requires taking a sufficient number of iterations.  The efficiency depends on how rapidly these correlations decay on successive iterations.  We investigate this by computing empirically a random variate its autocorrelation statistics as given in equation~\ref{equ_iter_autoCorr}.  We consider the fluid within a spherical domain with no-slip Dirichlet boundary conditions with a finite element discretization having 460,904 degrees of freedom.  The autocorrelation is considered for direct stochastic Gauss-Seidel iterations and stochastic multigrid iterations
counting both the number of iterations and the number of Gauss-Seidel visitations to individual finite element nodes.  The visitations count more closely reflects the computational expense in acheiving a given level of 
decorrelation with a given method.  We find for the stochastic Gauss-Seidel iterations that many more iterations are required to acheive decorrelated variates, see Figure~\ref{fig_decorr_times}.  In fact, to achieve a correlation between the random variates of less than $1\%$ requires on the order of $100$ Gauss-Seidel iterations.  In contrast, the stochastic multigrid method achieves a correlation less than $1\%$ in less than $10$ iterations.  To further characterise the performance of the stochastic iterative methods, we considered how the random variates decorrelate as the Gauss-Seidel smoother visits individual finite element nodes.  We find that standard Gauss-Seidel iterations achieve correlations between the random variates less than $1\%$ after $10^8$ vistiations.  In contrast, the stochastic multigrid method makes better use of Gauss-Seidel visitations through coarsening of the system and acheives correlations between the random variates less than $1\%$ after $10^6$ visitations (two orders of magnitude less).  These results show the significant improvement provided by using stochastic multigrid to sample random variates.  The differences between the stochastic sampling methods are expected to become even more significant when increasing the system size.

\newpage
\clearpage

\subsection{Hydrodynamic Coupling}
Many approaches can be used to couple the microstructure to the fluid including the Force Coupling Method~\cite{Maxey2001}, Discontinous Galerkin Finite Element Methods~\cite{Persson2013, Lew2008}, or the Immersed Boundary Method~\cite{Peskin2002,Bringley2008}.  The simplest of these is to couple the microstructure and fluid using the Immersed Boundary Method~\cite{Peskin2002,Bringley2008}.  This corresponds to the specific choice of coupling operators
\begin{eqnarray}
\Lambda_{\bX}(\bm{F}) & = & \bm{F}(\bm{x})\delta_a(\bm{x}-{\bX})\\
\Gamma \bm{u}         & = & \int_{\Omega}\bm{u}(\bm{x})\delta_a(\bm{x}-{\bX})~d\bm{x}.
\end{eqnarray}
The $\delta_a$ is the Peskin-$\delta$ function which is non-zero over 
the distance $a$~\cite{Peskin2002}.
To adopt this approach in the current finite element setting the operators 
are discretized by approximating the integral for $\Lambda$ using 
the finite element basis to obtain the numerical coupling operator 
$B_\mb{X}(\cdot)$.  The other coupling operator $\Gamma$ is approximated
by using the adjoint condition in equation~\ref{equ_discr_SELM_full} which yields the numerical 
operator $B_\mb{X}^T$, see Appendix~\ref{appendix_IB}.

\begin{figure}[h]
\centering
  \includegraphics[width=0.95\textwidth]{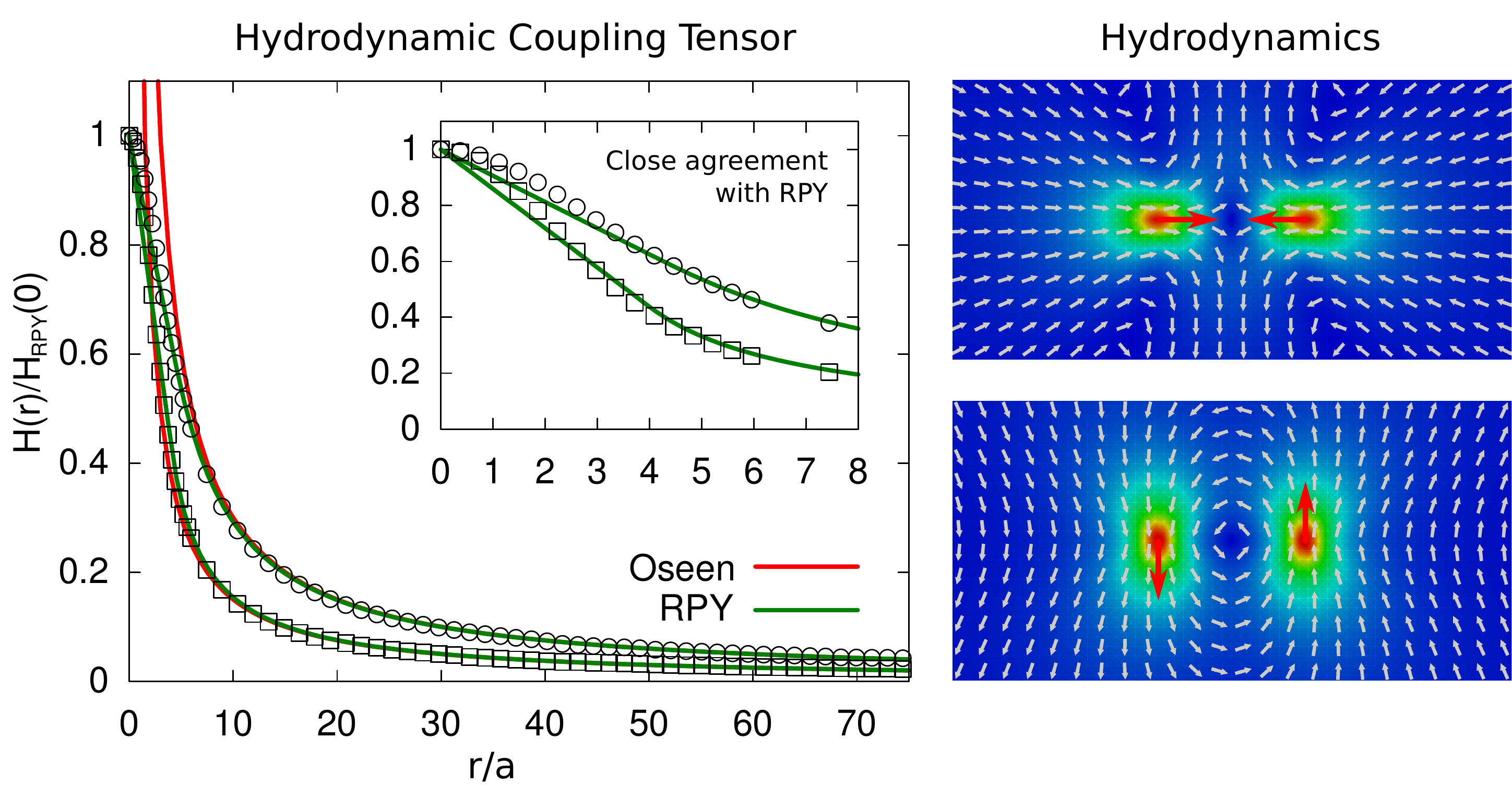}
\caption{Hydrodynamic Coupling.  We find the \textit{Immersed Boundary Coupling} results in hydrodynamic
coupling that closely resembles the Rotne-Prager-Yamakowa Tensor, see right panel~\cite{AtzbergerShear2009,Rotne1969,Yamakawa1970}.  
Both tensors in the near-field provide a regularization of the hydrodynamic coupling.  In the far-field both tensors 
give the same results as the Oseen tensor.  The hydrodynamic flow generated under the \textit{Immersed Boundary Coupling}
in response to forces acting on the two particles is shown on the right.}
\label{fig_HydroCouplingTensor}
\end{figure}

For this approximate approach to the physics of fluid-structure interaction, we investigate the hydrodynamic response when forces are applied to a pair of particles and their velocities.  This corresponds to averaging the fluid using 
$\Gamma$ and solving the steady-state Stokes equations when forces are spread to the fluid using $\Lambda$.   This response is characterized by the effective hydrodynamic coupling tensor $H_{SELM}$.  We find that the effective hydrodynamic coupling tensor $H_{SELM}$ very closely agrees with the Rotne-Prager-Yamakawa tensor $H_{RPY}$~\cite{Rotne1969,Yamakawa1970}, see Figure~\ref{fig_HydroCouplingTensor}.  This is in agreement with our prior work in the context of finite volume methods reported in~\cite{AtzbergerShear2009}.  These results establish the approximate way in which fluid-structure interactions are handled give the expected results in the far-field and in the near-field exhibit a well-characterized regularization.

\newpage
\clearpage

\subsection{Diffusion of a Particle in Harmonic Potential}

\begin{figure}[htpb]
  \centering
  \includegraphics[width=0.8\textwidth]{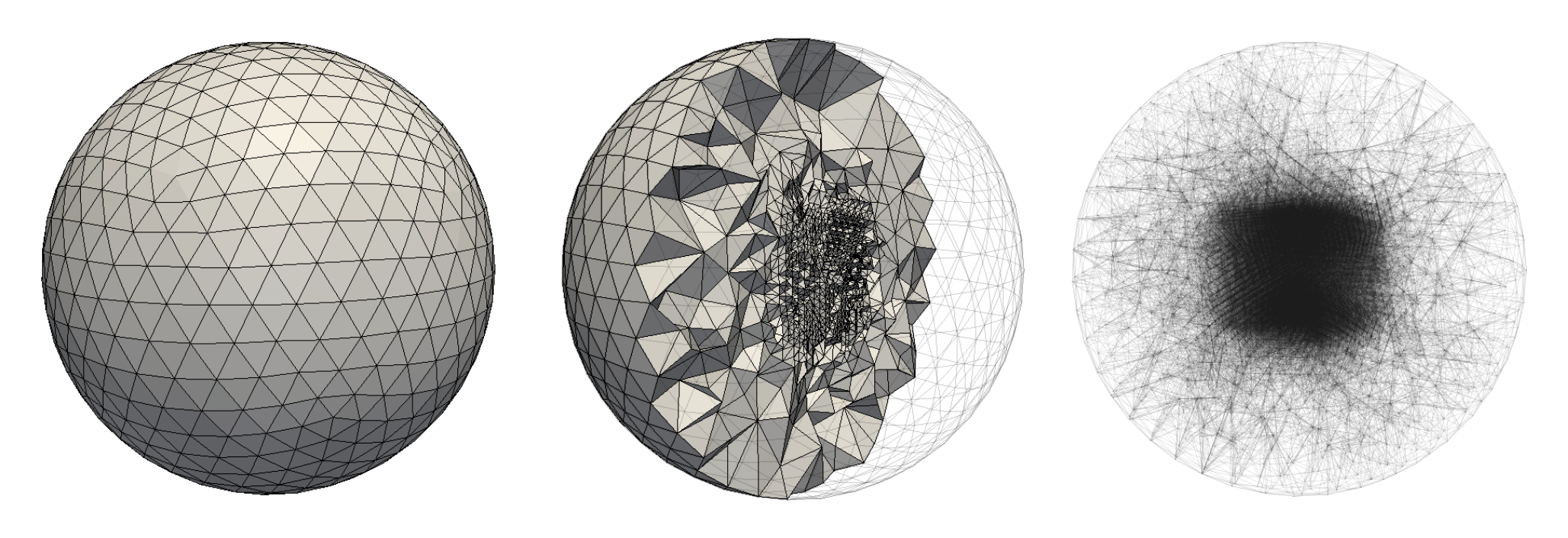}
  \caption[Spherical Domain and Adaptive Discretization.]{
Spherical domain discretized by a tetrahedralization with non-uniform spatial resolution.  The full spherical domain is 
shown on the left, a cross-section revealing the interior elements is shown in the middle, a wireframe representation of 
the elements is shown on the right.  This specific discretization was used in the
validation studies of a Brownian particle diffusing in a harmonic potential.  
Given how the harmonic potential is expected to confine particles toward the middle, a greater level of 
spatial refinement is used near the domain center.
}
  \label{fig:spherical_domain}
\end{figure}

\begin{figure}[h]
\centering
\includegraphics[width=0.32\textwidth]{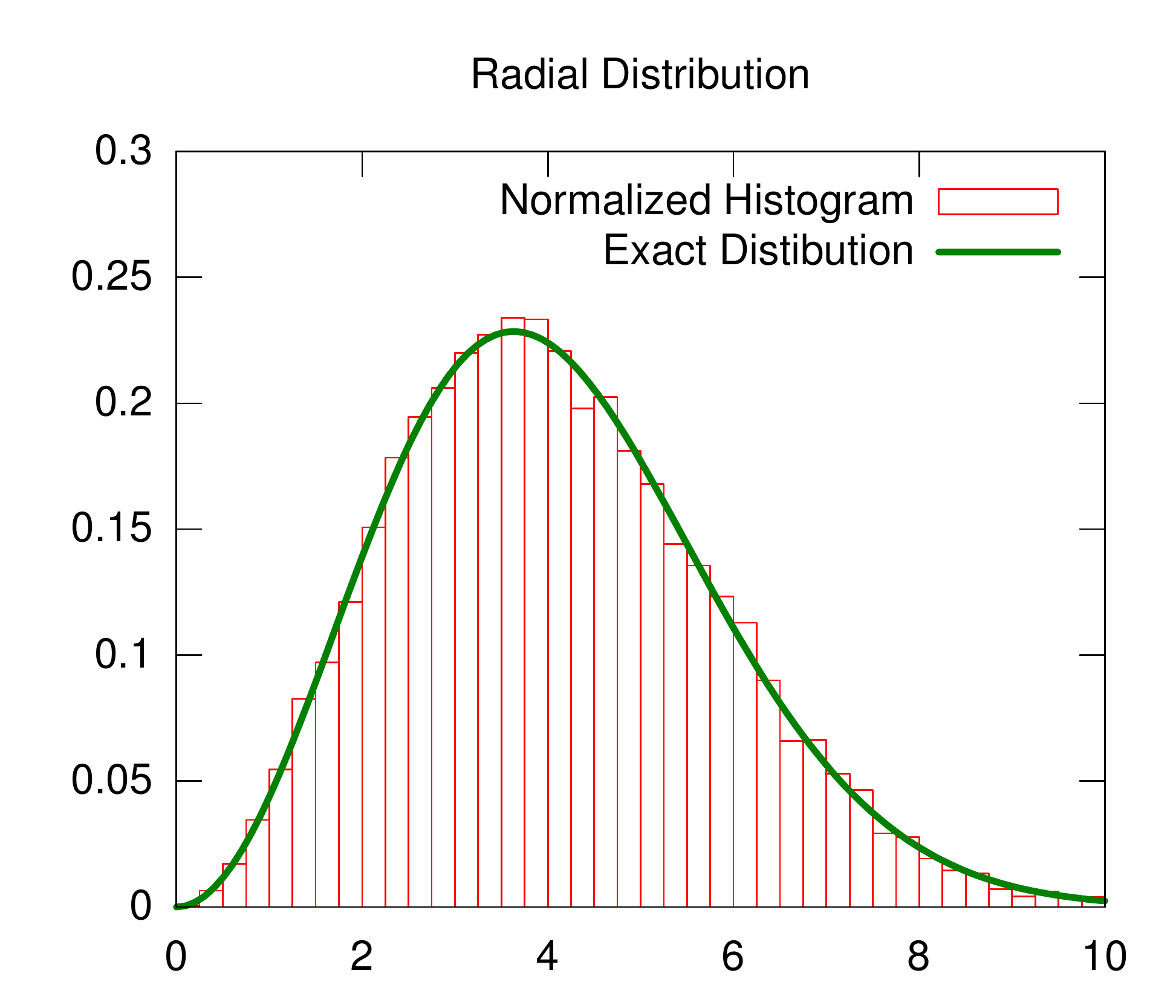}
\includegraphics[width=0.32\textwidth]{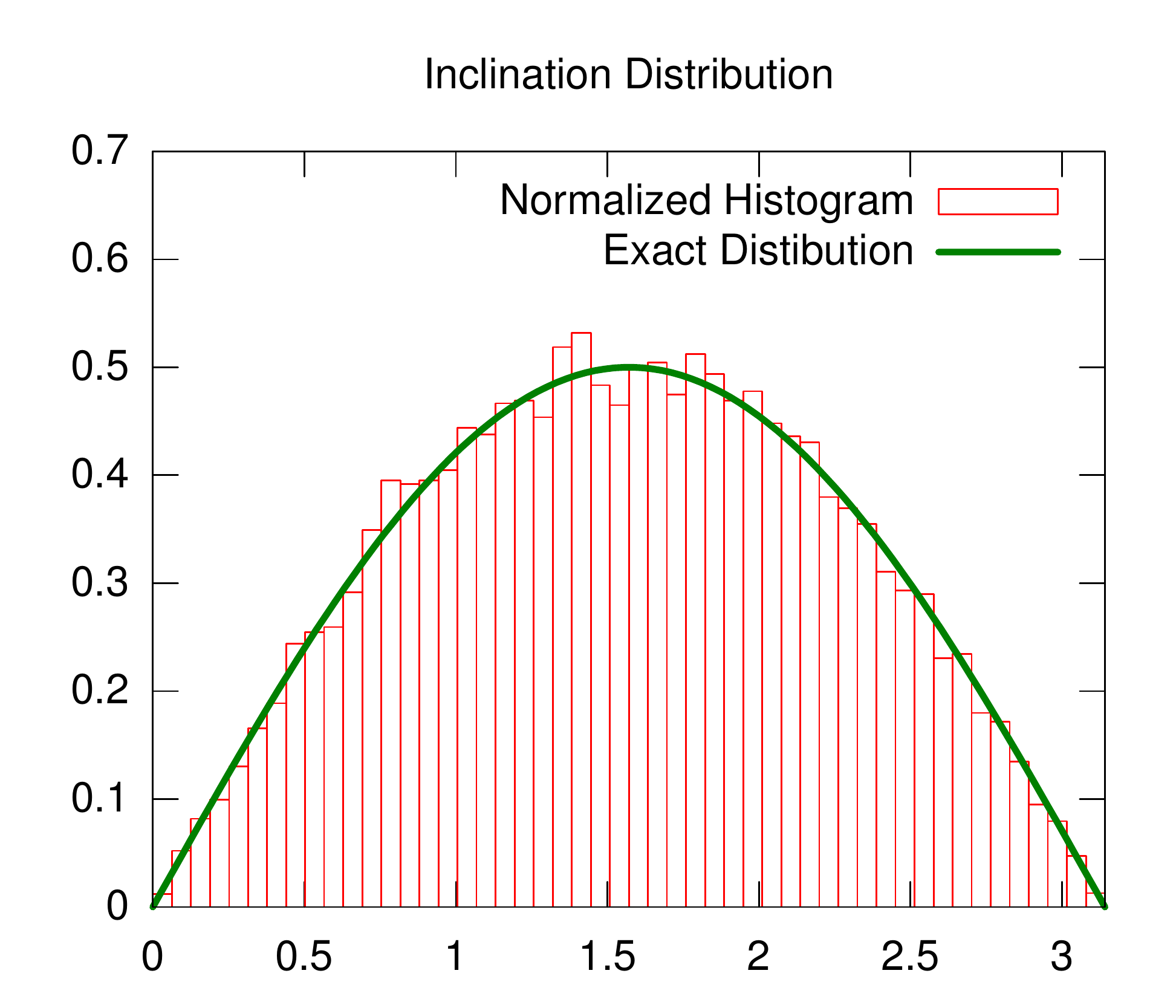}
\includegraphics[width=0.32\textwidth]{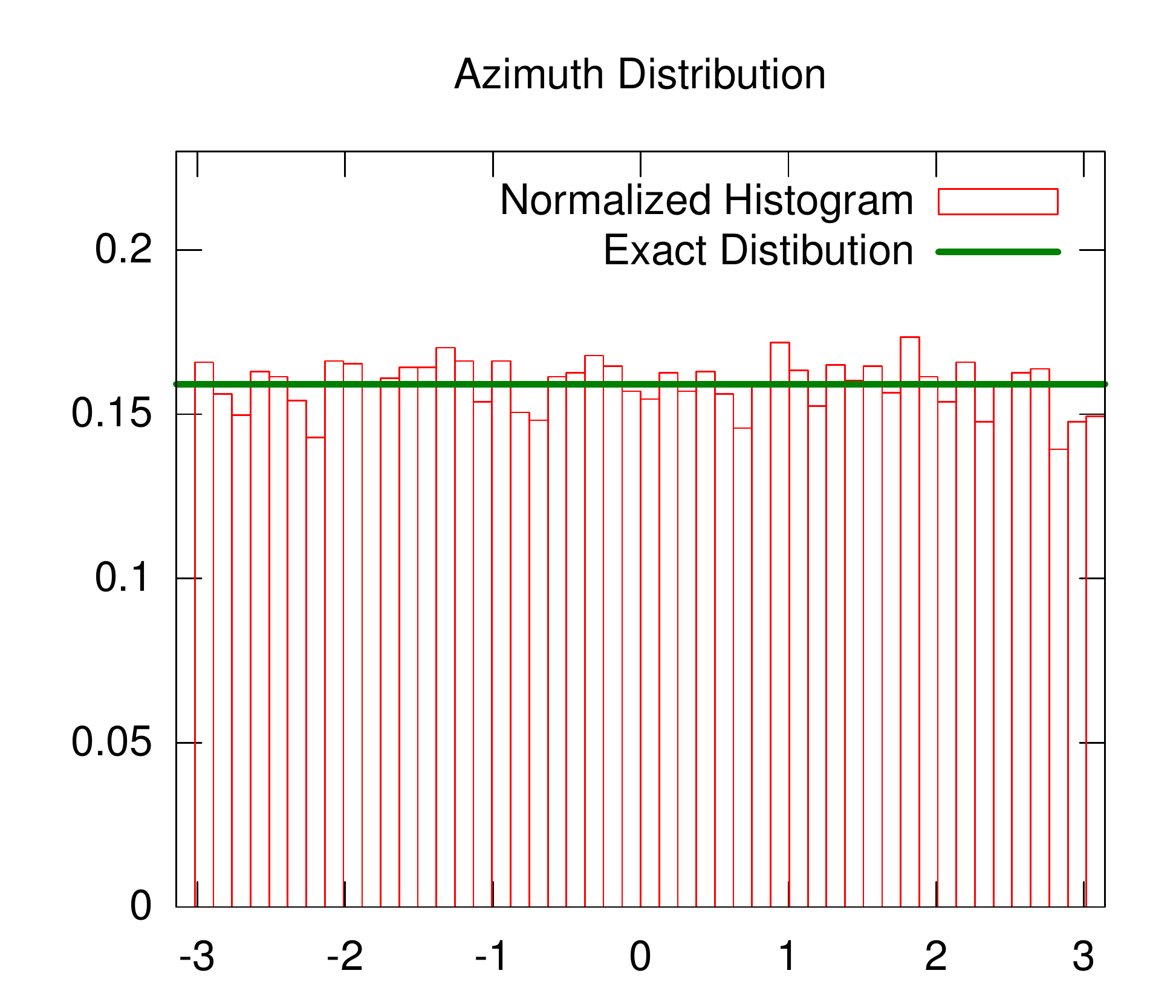}  \\
\includegraphics[width=0.32\textwidth]{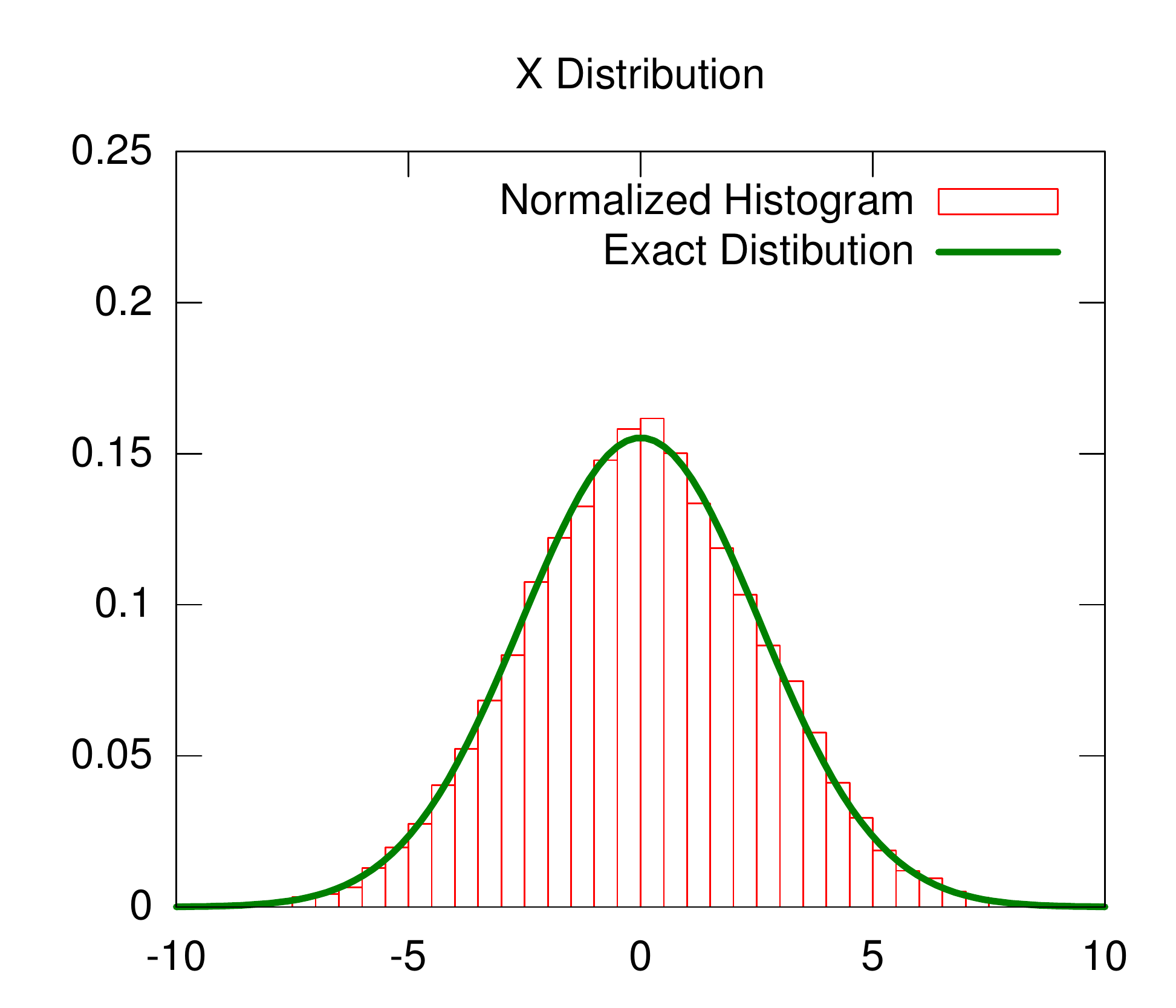}
\includegraphics[width=0.32\textwidth]{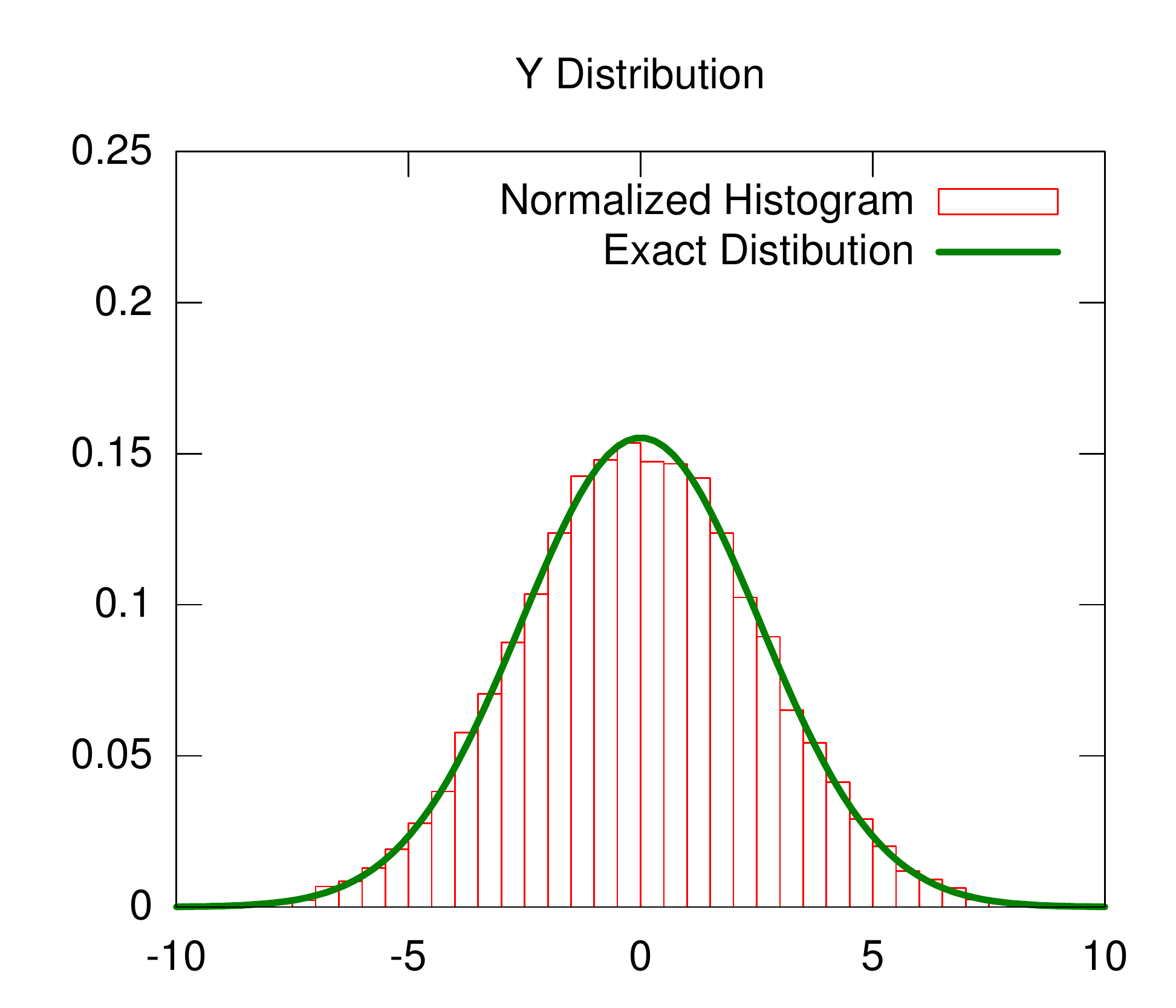}
\includegraphics[width=0.32\textwidth]{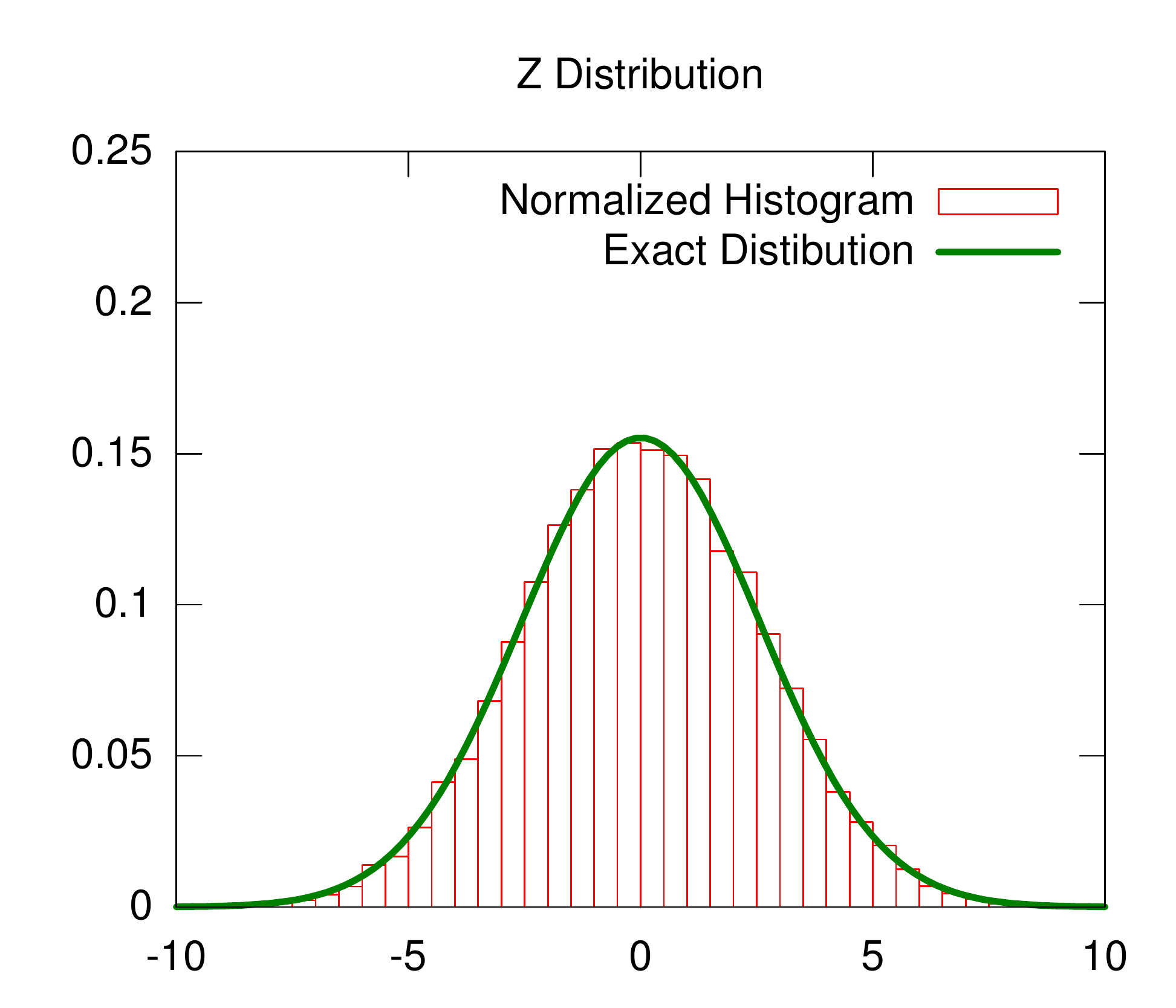}
\caption{Probability density of a particle's location when diffusing in a harmonic potential.  The probability distribution is shown
for the particle location when expressed in spherical coordinates $(r,\phi,\theta)$ (top) and Cartesian coordinates $(x,y,z)$ (bottom).  
The simulations results (red) show good agreement with the Gibbs-Boltzmann distribution (green).  }
\label{fig_GB_test}
\end{figure}

To investigate the fluctuations of the discretized fluid-structure system, we consider the simulation of diffusing particles that are subjected to forces from a harmonic potential that tether each particle to the origin.  From equilibrium statistical mechanics the probability distribution of the particle positions is predicted to be the Gibbs-Boltzmann distribution 
\begin{eqnarray}
\Psi(\mb{X}) = \frac{1}{Z} \exp\left(-\frac{E[\mb{X}]}{k_B{T}}\right), \hspace{1cm} E[X] = \frac{K}{2}\mb{X}^2.
\end{eqnarray}
The $E$ denotes the energy of a configuration, $K = 7.455504\times 10^{-1} \mbox{ag/ns$^2$}$ the Hookean spring stiffness, $T = 300 K$ the temperature,  and $Z$ denotes the partition function and normalizes the probability density~\cite{Reichl1998}.  The physical units we use are attograms (ag), nanoseconds (ns), nanometers (nm), and Kelvin (K).  The particles are taken to diffuse with an effective radius of 10nm 
on a spherical domain of radius 1000nm, see figure \ref{fig:spherical_domain}.  Samples were collected from 18 separate trajectories each of which has at least 1000 timesteps each.  The first 10 timesteps were discarded in each case.   We find that the computational simulations show very good agreement with the predictions of statistical mechanics, see Figure~\ref{fig_GB_test}.  These results show that the stochastic driving fields we have introduced and the stochastic iterative sampling methods provide appropriate fluctuations in the discrete system to account for the thermal fluctuations.

\section{Application}
\label{sec_application}

\begin{figure}[h]
\centering
\includegraphics[width=0.9\textwidth]{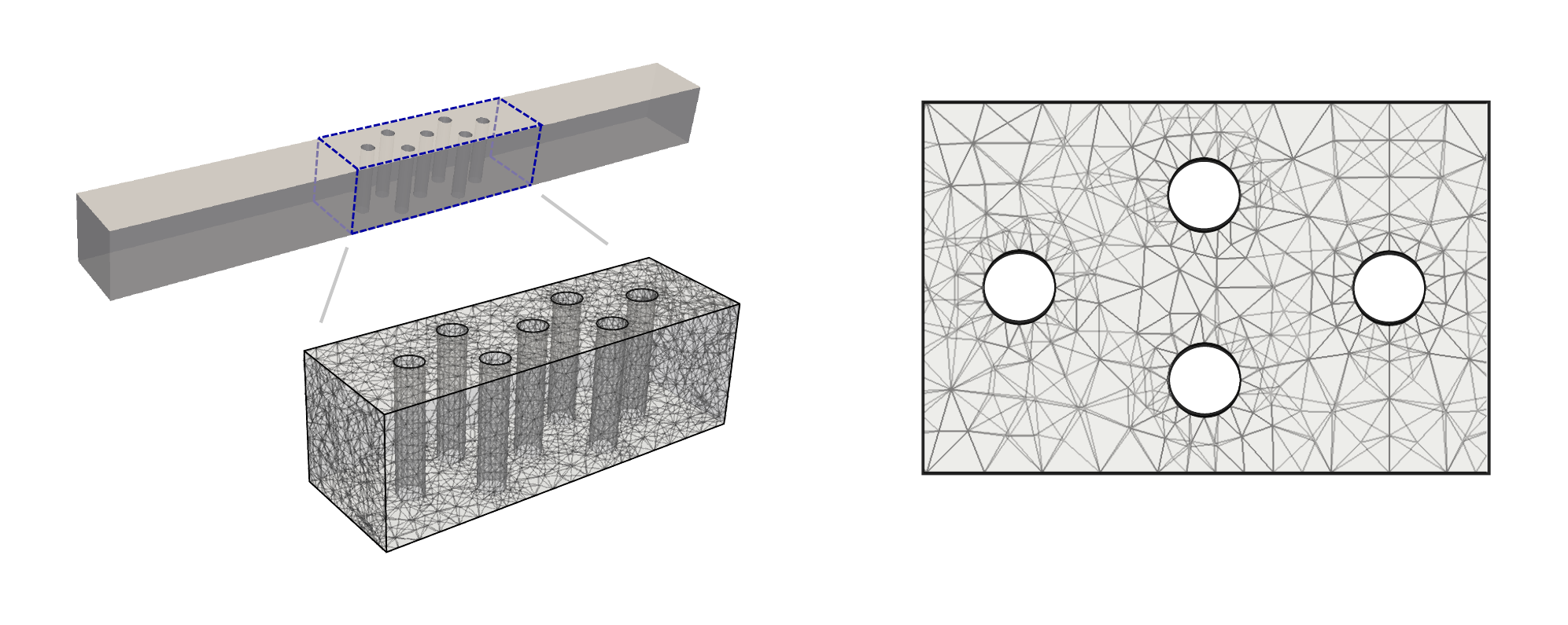}
\caption{Diffusion of particles in a domain with geometry motivated by microfluidic device (left).  The domain consists of a long channel with cylindrical posts obstructing flow near the center (right).  The channel side walls and cylindrical posts are treated as having no-slip boundary conditions.  The long-axis of the channel has periodic boundary conditions at the ends.  The domain is discretized using a tetrahedralization and the $\mathbb{P}_1$-MINI finite elements.  }
\label{fig_channel_schematic}
\end{figure}

\subsection{Diffusion of Particles within a Microfluidic Device Geometry}
We show how our methods can be used to capture effects on particle diffusion/mobility when 
confined within a microfluidic channel having a complex geometry.  We show how our methods can be used for hydrodynamic flows through post-like obstacles to capture the particle-wall hydrodynamic interactions and related correlation effects in the thermal fluctuations of diffusing particles.  The specific device geometry and our finite element discretization using the $\mathbb{P}_1$-MINI elements of Section~\ref{sec_element_choice} is shown in Figure~\ref{fig_channel_schematic}.  

\begin{figure}[htpb]
  \centering
\includegraphics[width=0.8\textwidth]{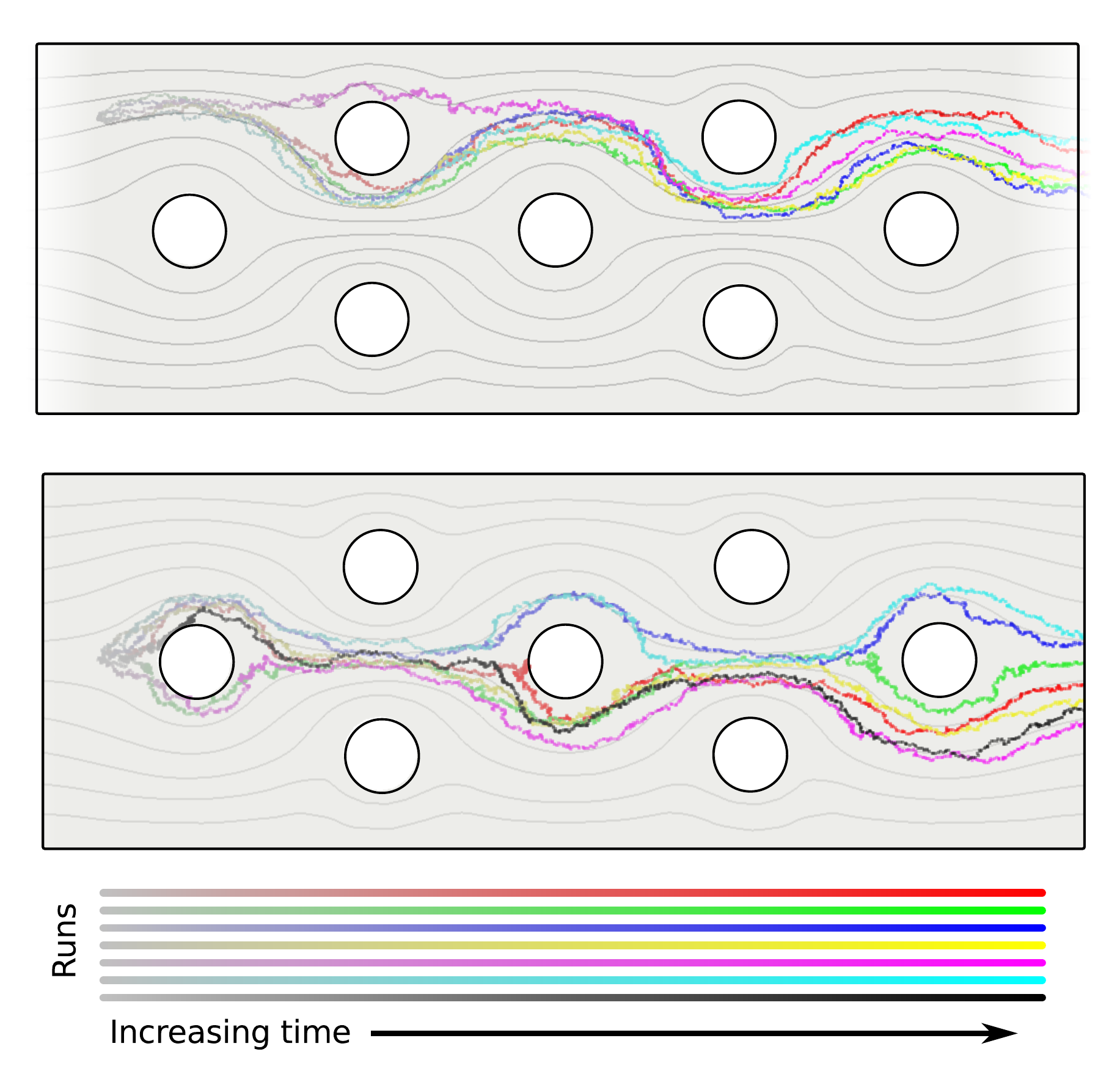}
  \caption[Drifting particle trajectories in a channel with obstructing posts.]{Diffusivity of particles started at select locations along a cross-section of the channel.  Particles are subject to hydrodynamic flow through the device channel from left to right.  The streamlines 
  show the influence of the channel walls / cylindrical posts and traceout how passive particles would be transported by the flow.  For particles subject to thermal fluctuations the diffusivity can causes particles to change streamlines.  For instance, the bottom magenta trajectory starts on a streamline going below the post but the diffusivity moves the particle to a streamline carrying it on the other side of the post.  In another example, the trajectory in black (bottom panel) shows a particle which migrates very little when it comes into close proximity to the leftmost channel post which reduces both the fluid flow and particle diffusivity.}
  \label{fig_diffusivity_vs_location}
\end{figure}

We consider the diffusion of particles in a regime of moderate flow through the device.  In this case, the main role of diffusion is in the directions lateral to the flow serving to change over time the effective streamline followed by a particle.  Given the geometry and small dimensions of the device relative to the particle size, significant hydrodynamic coupling can occur between the particle and walls.  The diffusivity may change significantly depending on the particular particle location within the channel.  To investigate this effect, we performed simulations by starting particles at several locations within the microfluidic channel.  These included (i) placing particles near the channel wall, (ii) placing particles near the cylindrical posts, and (iii) placing particles near the channel center.  Simulation trajectories of particles in these locations is shown in Figure~\ref{fig_diffusivity_vs_location}.  The trajectories starting near the wall and cylindrical posts seem to exhibit smaller fluctuations than when 
started near the channel center away from such boundaries.  To investigate this further, we consider in the absence of flow the particle mobility which is closely related to the location dependent particle diffusivity by
\begin{eqnarray}
D = 2k_B{T} M.
\end{eqnarray}
The particle mobility $M$ gives the velocity response $\mb{V}$ of a particle to an applied force $\mb{F}$
\begin{eqnarray}
\mb{V} & = & M \mb{F}. 
\end{eqnarray}

\begin{figure}[h]
\centering
\includegraphics[width=1.0\linewidth]{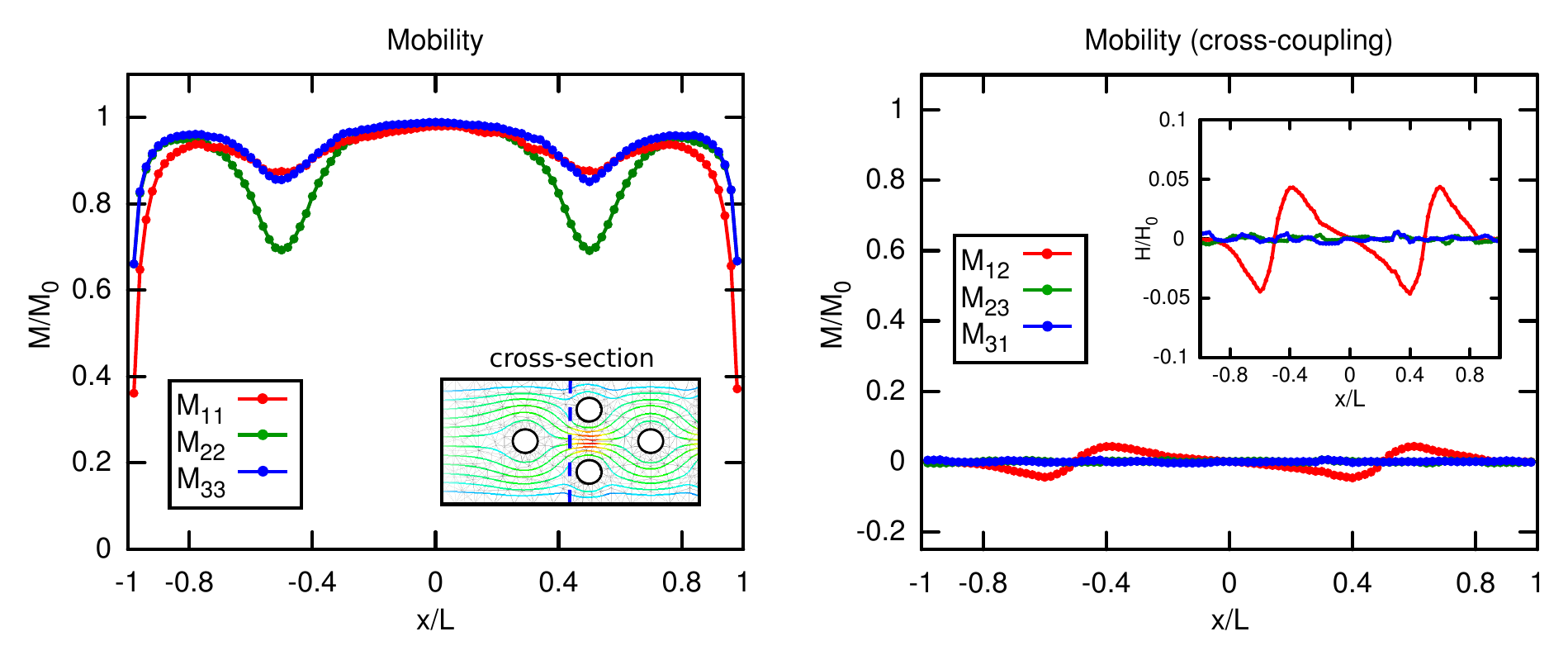}
\caption{Mobility dependence on the particle location within the channel.  The mobility is shown for a line intersecting the channel near the cylindrical posts (inset).  The mobility significantly drops to around $50\%$ near the channel walls then recovers but drops again by about $30\%$ when encountering the cylindrical posts. The region near the cylindrical posts is also found to induce some cross-coupling in the mobility components, particularly $M_{12}$.  These results show the potential for our computational methods to capture in the particle motions the effects of walls and related confinement effects.  The full three dimensional geometry is shown in Figure~\ref{fig_channel_schematic}}
\label{fig_mobility_vs_location}
\end{figure}

We consider how the mobility changes based on the particle location along a line of positions passing across the channel near to the cylindrical 
posts, shown in Figure \ref{fig_mobility_vs_location}.  As the particles become close to one of the cylindrical posts we expect a drop in the mobility.  This was found to be significant over the positions considered with a drop in mobility of around $30\%$ relative to locations away from the post.  While the most significant drop in mobility occurs in the direction for the particle to move toward the cylindrical post, we find the other directions also exhibit a non-negligible drop.  We also found cross-terms in the mobility which indicate that the cylindrical post geometry induces correlations between the particle motions in different directions, shown in Figure~\ref{fig_mobility_vs_location} on the right.  In particular, we find that the components in the direction from the particle to the post and in the lateral direction give non-negligible correlations.  This is in agreement with the 
streamlines that are found when there is fluid flow through the device, see inset in Figure~\ref{fig_mobility_vs_location}.  These results indicate that the diffusivity of particles can exhibit significantly different qualitative behaviors depending on the particle location within the device.  The results show the overall potential of our computational methods to capture both the thermal fluctuations and hydrodynamic coupling in a consistent manner within non-trivial device geometries.  While we demonstrated the methods for particles, our approach can also be used to simulate the elastic responses to flow of more complex spatially extended microstructures such as polymeric filaments or flexible membranes.   Overall, we expect our computational methods to be applicable quite broadly to the simulation of systems involving hydrodynamically coupled microstructures confined within domains with complex geometries and subject to thermal fluctuations.

\newpage
\clearpage

\section{Conclusions}
We have developed finite element discretizations for fluctuating hydrodynamic simulations of fluid-structure interactions subject to 
thermal fluctuations.  We introduced a discrete fluctuation-dissipation principle to obtain semi-discretizations of the 
stochastic equations.  Our approach accounts explicitly for the role of discretization errors on the thermal fluctuations 
so that the Gibbs-Boltzmann distribution is invariant exactly under the 
stochastic dynamics of the semi-discretized system.  To compute efficiently these required stochastic driving fields, 
we developed a Gibbs sampler based on stochastic iterations of multigrid.   The computational methods are expected 
to extend significantly the applications that can be treated with fluctuating hydrodynamic approaches.  The computational
methods allow for spatially adaptive resolution of the mechanics and the treatment of complex geometries often 
relevant in application.

\section{Acknowledgments}
The authors would like to acknowledge support from DOE ASCR CM4.  The author P.J.A. acknowledges 
support from research grant NSF CAREER DMS-0956210 and W.M. Keck Foundation.  The Trilinos packagess 
ML and Epetra were used for the application simulations.  The authors thank Alexander Roma, 
Boyce Griffith, Mike Parks, and Micheal Minion for helpful suggestions.  

\bibliography{manuscript}   

\appendix

\section{Constructing $B_{\mathbf{X}}$ for Immersed Boundary Method Coupling}
\label{appendix_IB}
To approximate the coupling operator $\Gamma[{\mathbf{X}}] \bm{u} = \int_{\Omega}\mb{u}(\mb{x})\delta_a(\bm{x}-{\bX})d\bm{x}$, we use in the finite element method the semi-discrete operator
\begin{eqnarray}
  B_{\bX}\mb{U} &=& \sum_j\mb{U}_j\int_{\Omega}\phi_j(x)\delta_a(\bm{x}-\bX)~d\bm{x}.
\end{eqnarray}
We use for $\delta_a$ the Peskin $\delta$-function approximated by the cosine described in~\cite{Peskin2002}:
\begin{displaymath}
  \delta_a(x_1,x_2,x_3) = \frac{1}{a^3}\phi\left(\frac{x_1}{a}\right)\phi\left(\frac{x_2}{a}\right)\phi\left(\frac{x_3}{a}\right),
\end{displaymath}
where
\begin{displaymath}
  \phi(r) =
  \begin{cases}
    \frac{1}{4}\left(1+\cos\left(\frac{\pi r}{2}\right)\right), & |r| \leq 2,\\
    0, & \text{otherwise}
  \end{cases}.
\end{displaymath}
We assume an immersed body $\bX(\mb{q})$ parameterized by $\mb{q}$ is discretized into a finite number of control points $\bX_i$.  Thus, the \emph{semi}-discrete operator $B_{\bX}$ can be written as the \emph{fully} discrete operator with entries
\begin{displaymath}
  (B_{\bX})_{ij} = \int_{\Omega}\phi_j(\bm{x})\delta_a(\bm{x}-\bX_i)~d\bm{x}.
\end{displaymath}
We approximate these entries using on each tetrahedron of the finite element mesh the degree three Gaussian quadrature ($M=4$) described in~\cite{Yu1984}.  Other choices of operators can also be used to couple the fluid and microstructures~\cite{AtzbergerSELM2011}.
\end{document}